# Modeling Multi-Magnet Networks Interacting Via Spin Currents

Srikant Srinivasan[1,2], Vinh Diep[1], Behtash Behin-Aein[1,3], Angik Sarkar[1] and Supriyo Datta[1],
[1]*School of Electrical & Computer Engineering, Purdue University, West Lafayette, IN-47907*
[2]*Dept. of Mat. Sci. & Eng., Iowa State University, Ames, IA-50011,* [3]*Globalfoundries Inc.*

# 1. Introduction

## 1.1 Information processing using Spin currents and Nanomagnets

A promising candidate in the quest for alternatives to charge-based transistors [1, 2] has been a broad class of devices (for example: [3-6]) that propose to implement information processing using spin currents and nanomagnets. The significant experimental advances of the last few decades in dealing with the interaction of spin currents and nanomagnets at the device level has allowed envisioning large scale circuits based on these different proposals. In general, these experiments can be grouped under one of two distinct physical phenomena: (1) Injection of spin currents by magnets into semiconducting or metallic channels and the transport of spin currents within these channels[7-9] (usually laterally grown structures) and (2) Spin torque switching of magnetization [10-12] by injecting spin currents into a magnet (typically seen in vertically grown structures). In recent work (Behin-Aein, Datta [3], Srinivasan, Sarkar [13]) we have suggested how these two phenomena can be combined to form the basis of an All Spin Logic (ASL) scheme, which can be used to build computational logic blocks in a manner reminiscent of CMOS.

*The primary purpose of this chapter* is to bring to light, a flexible and powerful spin-transport/ magnetization-dynamics framework that we constructed to describe spin-magnet systems in general, and which was instrumental in modeling ASL [13, 14] for accurate switching behavior, energy-delay products, scaling trends and finding novel mechanisms of inbuilt directionality. Here, in the first part of this introduction, we will briefly review – in the context of spin-based information processing – how ASL represents a significant milestone in the development of realistic all-spin device/ circuit implementation.

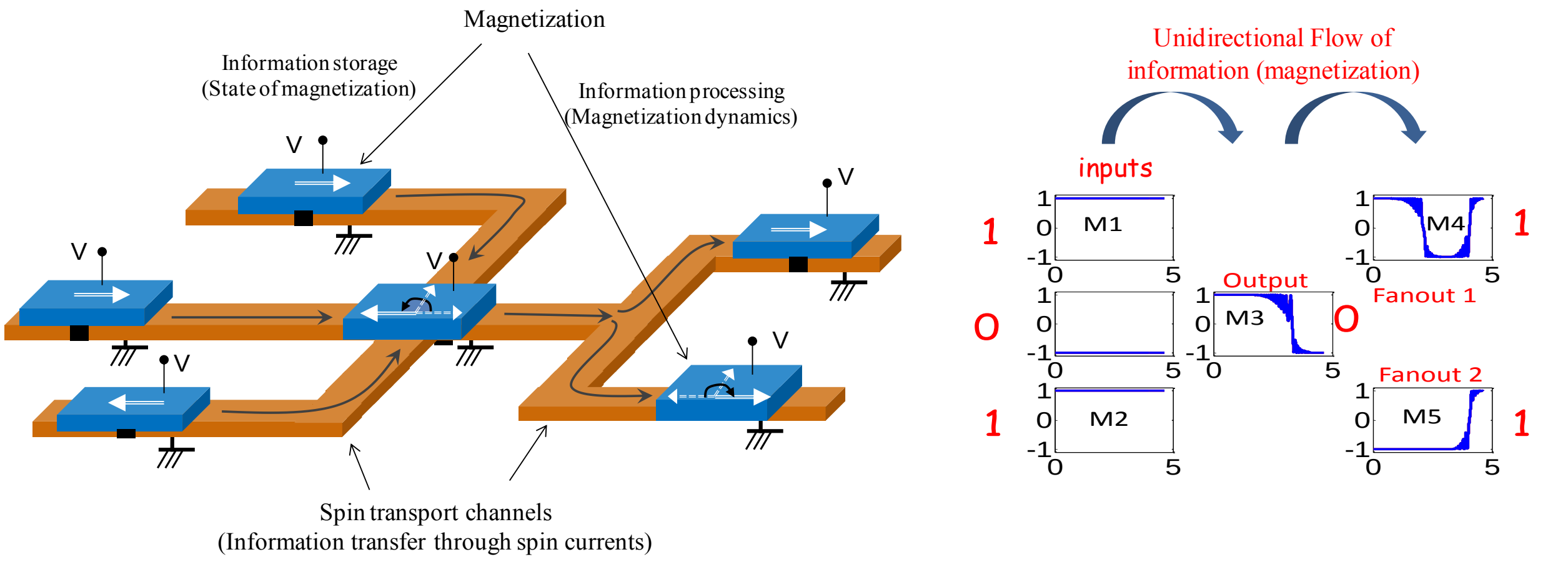

Fig. 1 – A computational circuit with spin currents and nanomagnets: The magnets receive input information in the form of spin currents via a spin transport channel. Information is processed by the switching action of the magnetization and the processed information is transmitted as spin current to the next stage. This particular layout represents a NAND logic gate (discussed in section 3). The simulated behavior of the magnetization is shown for a particular set of inputs.

Consider a typical ASL computational block shown in Fig. 1, comprising nanomagnets and spin transport channels. Each nanomagnet in this circuit can be thought of as a reservoir of similarly oriented spins. The magnet can be suitably engineered so that these spins collectively point along one particular direction, commonly referred to as the 'easy axis' of the magnet. The resulting magnetization pointing one way or the opposite along the easy axis now provides a natural representation for digital information, '0' and '1'. Information processing is achieved by making the magnetization toggle between '1' and '0' based on the spin current input to the magnets through the spin transport channels.

The operational principle of the individual ASL unit is quite similar to the non-local spin transfer torque (NLSTT) phenomenon shown in Fig. 2. In NLSTT [Fig. 2(a)] a charge current flowing to ground from a magnetic contact on the left (input), gives rise to a spin current to the right, outside the path of charge current. This "non-local" spin current has been shown to be capable of flipping the magnetization of a second magnet (output) on the right hand side [15, 16]. Thus, information is read from the left magnet by translation of the input terminal voltage into a spin current $I_S$, which transmits to the right and is written onto the second magnet (through spin-torque switching). Such information transfer from one nanomagnet to the next as a process of Read (R) followed by a Write (W) is discussed in Ref.[17]. We note here that similar information transfer can also be achieved in the "local" configuration, wherein charge current (and accompanying spin current) is directly driven from the input to the output magnet. However, the non-local configuration makes it easy to see the decoupled nature of charge and spin currents. For positive voltages $V_{in}$ the spin current has a sign opposite to that of the *input* magnet and the overall structure in Fig. 2(a) functions as an inverter.

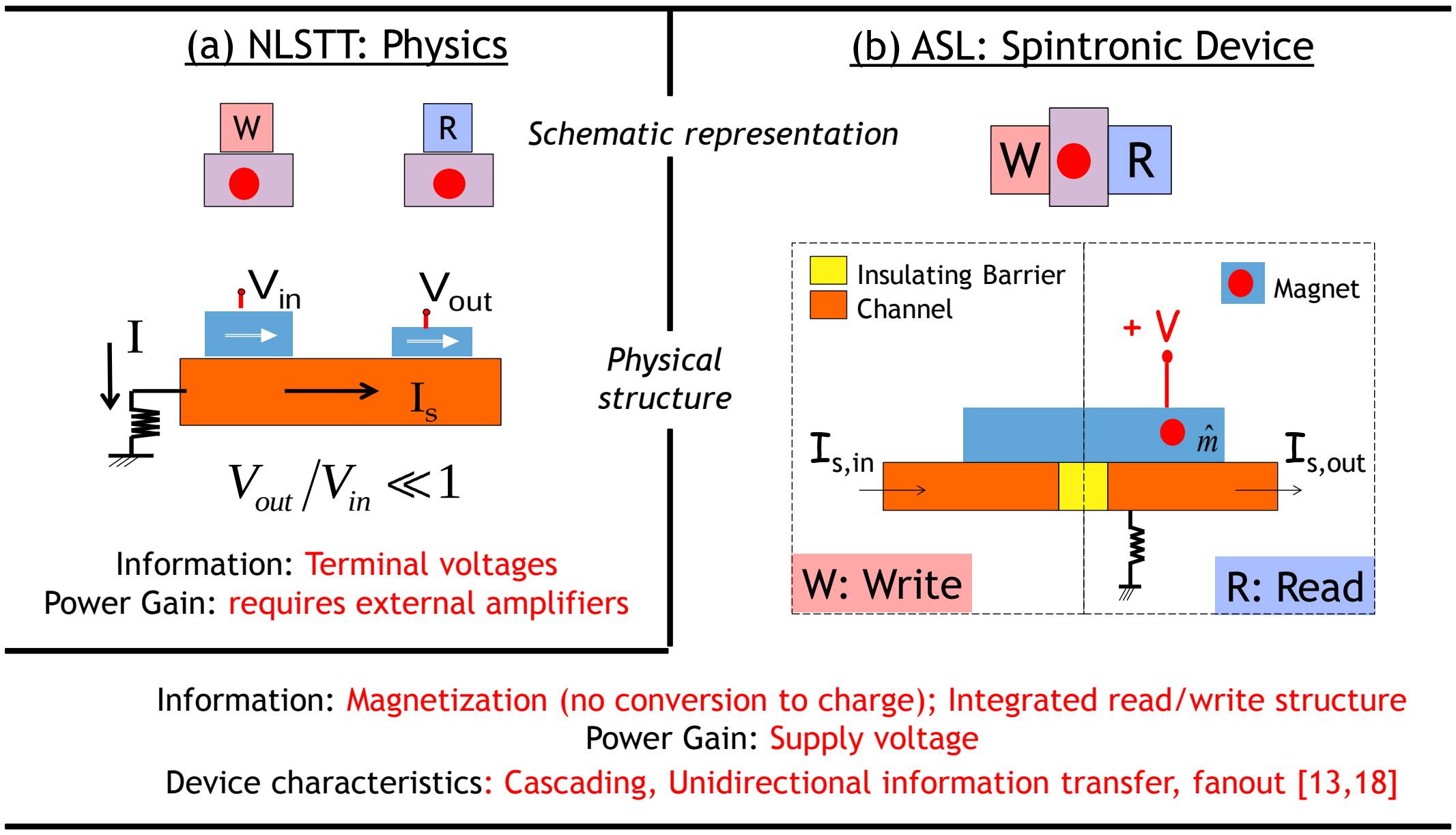

Fig.2 – (a) Non-local spin transfer torque (NLSTT) phenomena can be viewed as a read (R) followed by a write (W). However, as discussed in the text, not all *Read-Write* units can be cascaded to form large scale circuits. (b) The basic ASL *Read-Write* unit that incorporates additional device characteristics allowing circuits such as in Figs. (1) and (4).

It is important to note, however, that while *Reads* followed by *Writes* (perhaps after some processing) comprise the essence of logic, the mere availability of a *Read* and a *Write* mechanism in NLSTT (or a number of the other spin-transfer torque structures such as the popular magnetic tunnel junction) does not allow one to perform logic functions using just several of these devices together. This is because, in NLSTT, the measured quantity is a non-local voltage at the write terminal, several orders of magnitude smaller than the voltage at the read terminal. This ‘output’ voltage is found to be insufficient to drive subsequent stages unless it is interfaced through external amplifiers and/or sophisticated clocking mechanisms, requiring additional CMOS circuitry. This is what we could broadly refer to as “CMOS – dependent logic”, i.e, requiring transistor intervention at every stage. But, if we wish to develop a “self-contained logic” scheme that would allow us to interconnect hundreds of *Write-Read* (W-R) units without the use of any clocks or intervening CMOS circuitry, then it is important to design individual W-R units to have gain and directivity, properties that come naturally in transistor-based circuits, but not so easily with magnets. The ASL device [Fig. 2(b)] satisfies these additional device and circuit requirements because it incorporates a lot more than the basic NLSTT functionality.

In ASL, the external circuitry is avoided by designing W-R units with a transistor-like directionality: the magnet is controlled more effectively by the input spin current than by the output spin current, just as the channel of a Field Effect transistor is controlled more by the gate voltage than by the drain voltage. Every magnet has an insulating barrier beneath it (such as due to oxide deposition or a physical cut or doping etc), which allows it to interact separately with the preceding and succeeding stages through a non-magnetic channel. Ideally, for unidirectional information flow, we want one side of each magnet to behave as the *Write*, where it can receive information, and the other side to behave as the *Read* from where it can pass it on. Thus the device behaves as a self-contained current driven switch that does not require any intermediate charge – based conversion. The supply voltage in this case only serves to provide ‘power gain’ so that the *Read* side can drive subsequent stages. Having the same supply voltage on all the ASL devices makes them function in the non-local configuration while having different supply voltages makes them function in the local configuration. Much of our work on ASL has been focused on designing W-R units of the type shown in Fig. 2(b) and establishing [Fig. 3(a)] that they exhibit sufficient gain and unidirectionality to allow large scale circuit implementation [13, 18, 19] analogous to CMOS.

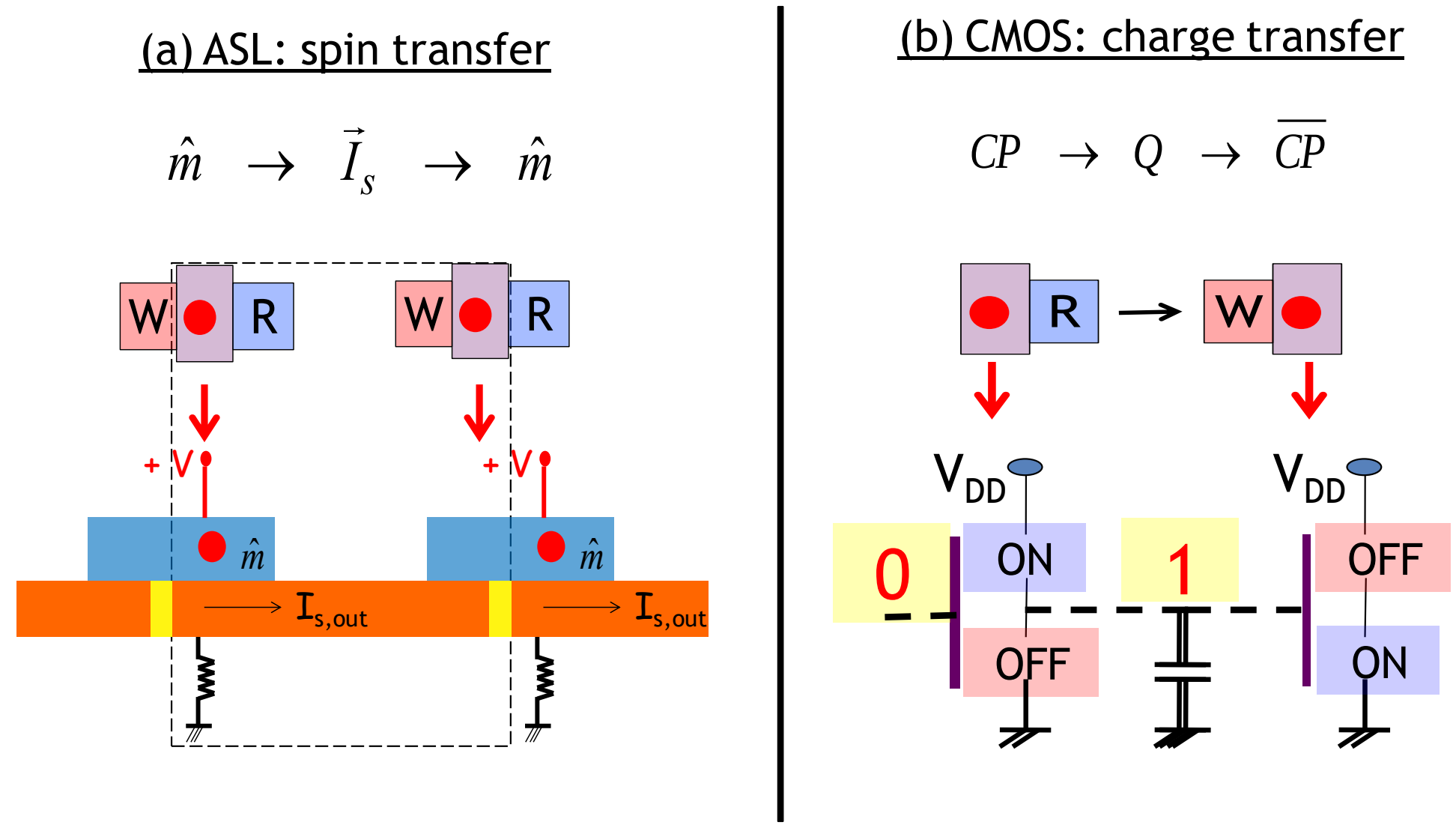

Fig. 3 – (a) Interconnection of basic W-R unit for ASL does not require any external amplifiers or clocks. The information is entirely in the spin domain decoupled from the supply voltage, which only powers the circuit. (b) A series of CMOS inverters can also be viewed as a sequence of reads and writes (Adapted from [17]).

Indeed one could view a series of standard CMOS inverters [Fig. 3(b)] as a sequence of *Reads* and *Writes* as well. We could say that each complementary pair (CP) plays the role of a magnet. The state of the CP is read by transforming it to charge on a capacitor – the gate capacitor of the next stage being charged by $V_{DD}$ or discharged by ground, through the CP - which is then written onto the next CP through the gate voltage. One advantage of magnets is that they are "non-volatile". Unlike the CP which loses all information once the power supply is removed, information stored in a magnet is "non-volatile". It is also a natural digital spin capacitor making it particularly well-suited for digital and neuromorphic circuits [3, 19, 20].

A signature result that distinguishes an inverter unit having gain from a passive one is a ring oscillator comprising an odd number of W-R inverters connected in a ring as shown in Fig. 4. Each unit, being an inverter, tries to switch the following unit in a direction opposite its own. With an odd number of inverters in the ring there is no overall stable state. If the spin current from each unit exceeds the threshold value needed to switch the following unit, then the z-component of the magnetization ($m_z$) of each unit exhibits continuous stable oscillations as shown. This behavior, well-known for CMOS inverters, is really quite surprising in the context of magnets. One does not expect three identical interconnected magnets powered by a constant d.c supply voltage to give rise to such controlled and predictable oscillations. This is made possible by the two key characteristics we mentioned, namely, gain and directivity: each unit is capable of switching the following unit without itself getting affected in turn.

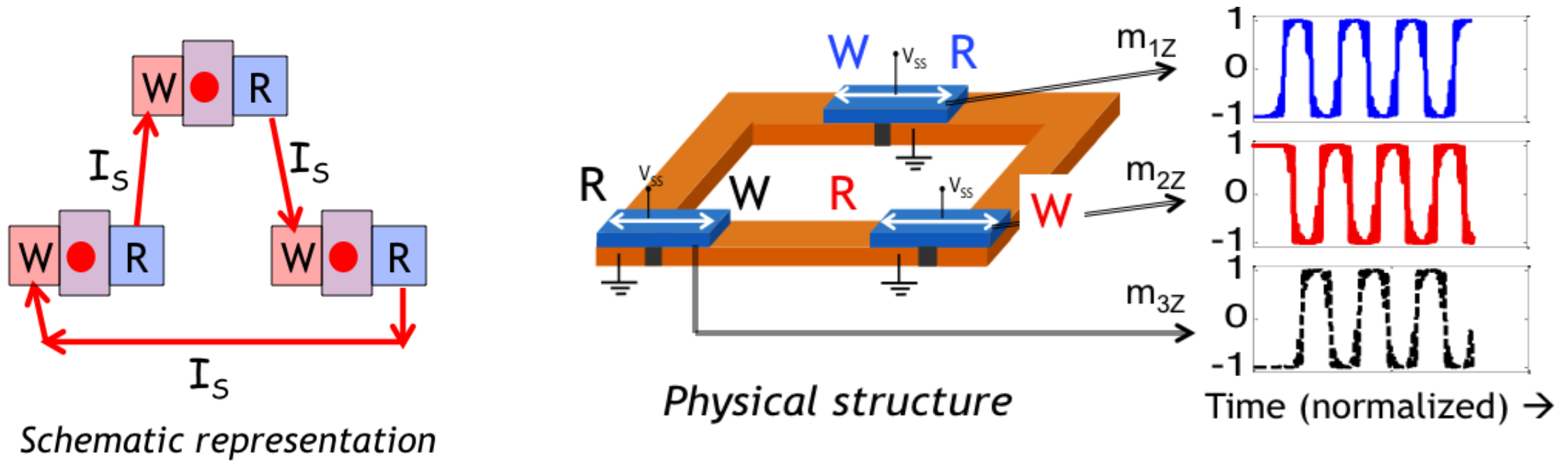

Fig. 4 – An odd number of inverters connected in the form of a ring comprises a ring oscillator where each unit periodically switches the following unit. Gain and directionality of individual W-R units are essential for the functioning of a ring oscillator.

## 1.2 A coupled spin-transport/ magnetization-dynamics model for spin-based device and circuit design

As mentioned earlier, to analyze spin-magnet logic circuits in general, we have developed an overall simulation framework (Fig. 5) simultaneously capturing spin – transport as well as magnetization dynamics, which is broadly useful beyond ASL. Indeed the primary purpose of this chapter is not to describe ASL circuits, which have already been adequately discussed in our earlier publications; rather it is to describe in detail the overall approach we have developed for the analysis of spin-magnet circuits and how it was benchmarked [14] against available data on spin – torque experiments. As can be noted from Fig. 5, the overall simulation framework contains two individual components –

- a circuit model for non-collinear spin transport (pertaining to the transfer of spin information between magnets) and

- a description of magnetization dynamics (pertaining to the mechanism of information processing and storage within the magnet).

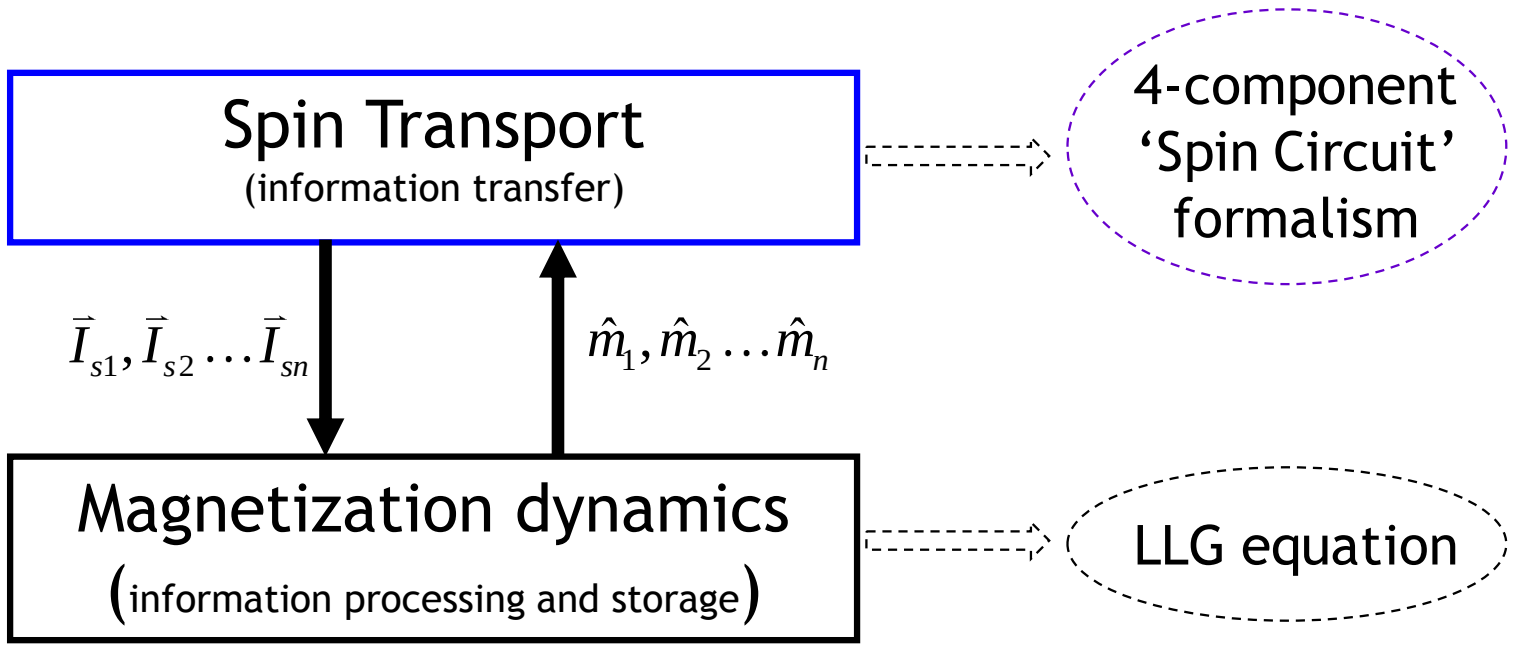

Fig. 5 – A coupled spin-transport/ magnetization-dynamics model for spin-magnet circuits. The magnetization dynamics relates to how (spin) information is processed and stored in the nanomagnets. This is described by the standard Landau-Lifshitz-Gilbert equation. The "spin transport" component models how information is transferred among the nanomagnets. We introduce a novel 4-component spin-circuit formalism to describe this part.

In Section 2 we address the *spin-transport component* using a lumped "4-component spin-circuit formalism" that we have developed, based on the work of the Bauer group [21-23], to describe the interaction of non-collinear magnets (required for modeling spin torque). This model computes 4-component currents and voltages at every node of a 'circuit'. Each nodal quantity has 4 components: one for the charge information and three components for the spin information corresponding to the x, y, and z directions. The use of a lumped element representation allows one to conveniently construct circuits in a modular fashion starting from basic device elements, quite similar to the SPICE modeling widely used for CMOS circuits. This approach to our spin-circuit model has been employed for modeling a broad spectrum of spintronic devices ranging from analysis of local spin valve structures [24] to domain wall propagation [25]. In addition, the use of a lumped circuit model facilitates insights into the working of complicated device geometries by straightforward transformation of such circuits into analytical expressions. This will be shown in section 2 with the example of deducing an expression for magnetoresistance of a non-local spin valve leading to the well-established result [26] for such structures.

For modeling the *magnetization dynamics*, we use the standard Landau-Lifshitz-Gilbert (LLG) equation with the Slonczewski [27] and field-like terms included for spin torque. Section 3 describes how this LLG model is coupled with the spin transport model to analyze existing experiments [15] and spin-magnet circuits in general.

We include MATLAB codes in the Appendix to facilitate a "hands-on" understanding of our model and hope it will enable interested readers to conveniently analyze their own experiments, develop a deeper insight into ASL or come up with their own creative designs.

# 2. Circuit representation of spin transport

As noted earlier the "spin transport" component pertains to modeling how information is transferred from one magnet to the next through spin transport channels. The modeling technique used for this purpose is the 4-component lumped pi-network model for spin transport, which is accurate in the linear and diffusive regime of transport. The use of lumped circuit elements in the model naturally lends itself towards simulating large scale circuits involving non-collinear magnets [Fig. 6 (a)]. Once the basic circuit elements are in place, the process of translation from arbitrary physical layouts [Fig. 6 (a)] to "simulatable" circuit constructs [Fig. 6 (b)] can literally be automated. The rest of this section is an elaboration of this process flow.

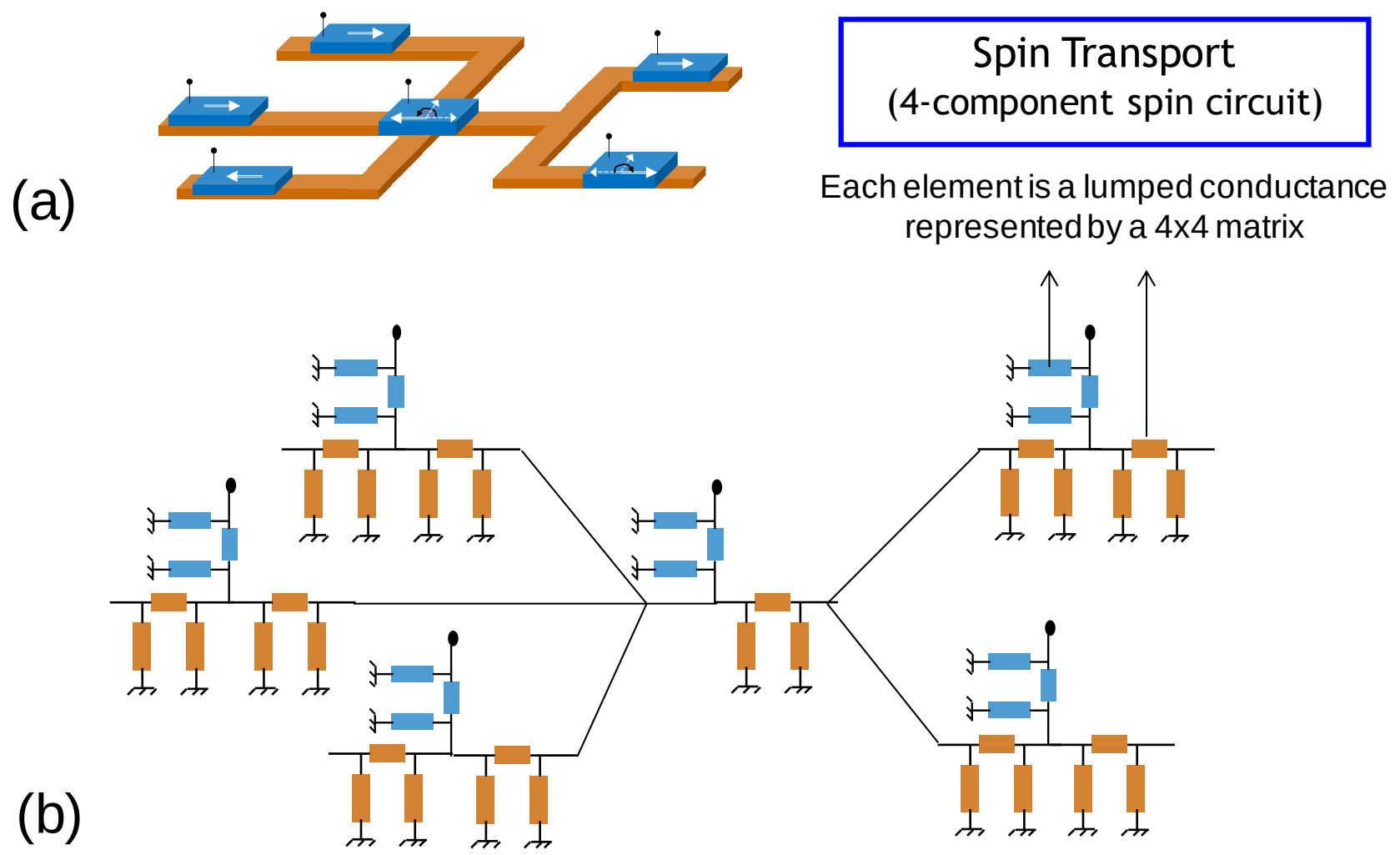

Fig.6 – Spin transport modeling using lumped circuit elements (a) The spin based computational circuit shown in Fig. 1, and (b) the equivalent spin circuit respresentation using 4x4 lumped pi-conductance matrices derived further along this chapter.

From a modeling perspective, it should be recognized that any physical structure designed for spin injection and spin transport can usually be resolved into three regions [Fig. 7]: (a) A non-magnetic channel that carries the spin current, (b) a magnet, which acts as a source of spin polarized carriers and (c) optionally an additional interface region which enhances the injection of spins from the magnet into the channel. Once we have a reliable lumped element representation for each of these regions it is easy to model large scale circuits since they are basically a combination of these three blocks. We will briefly review the existing theory for diffusive spin transport and show how it leads to a lumped circuit model.

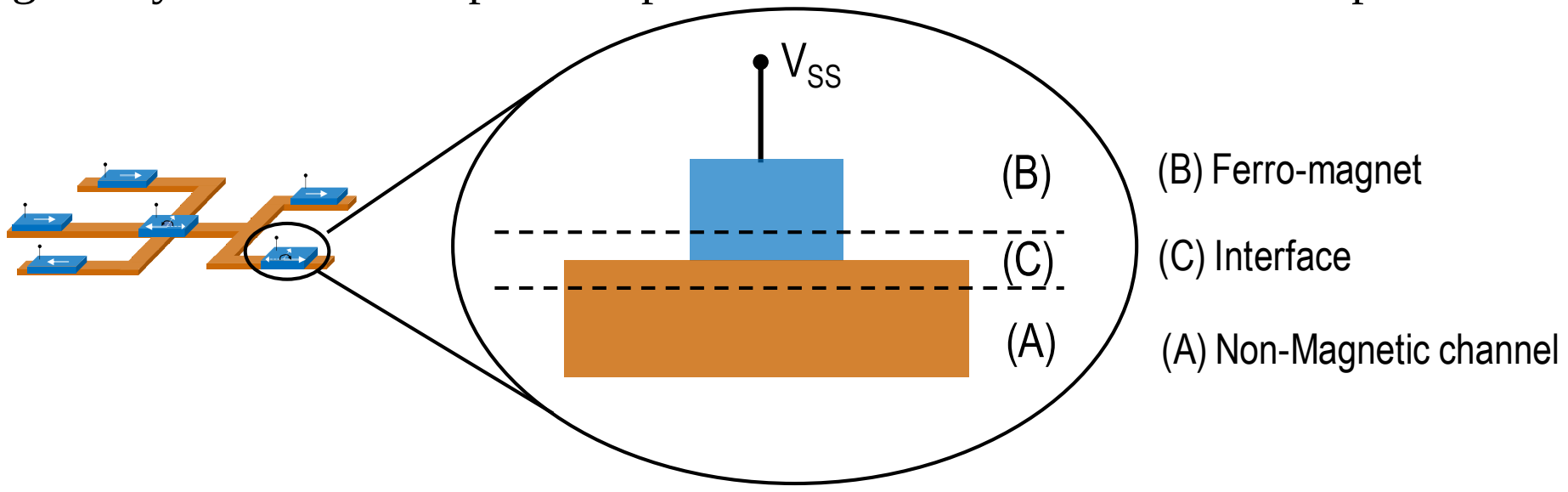

Fig. 7 – The three regions of any circuit involving magnets interacting via spin currents.

## 2.1 Spin diffusion equations and the distributed spin-circuit model

At the outset it is important to keep in mind that, depending on the regime of operation and types of physical mechanisms involved, there are different techniques available for modeling spin-magnet systems. An extremely low temperature experiment with correlation effects or involving high spin orbit coupling would require a full-quantum transport formalism like the Non-Equilibrium Greens function [28]or scattering theory [21] to describe the current flow. On the other hand, due to practical operational constraints, spin based computing on a large scale understandably involves room temperature operation and the transport regime in these systems is closer to diffusive, which is captured by the spin diffusion equations [29, 30].

We will now describe what we could call a "spin-circuit" approach, which is a distributed transmission line representation leading to the spin diffusion equations. The idea behind these equations is that, in spin diffusive channels, one can conceptually think of the up-spins and down-spins being transported through separate conducting channels [31], intermittently connected to each other through a spin flip conductor [Fig. 8]. Usually, when calculating quantities such as density of states or current flow in non-magnetic materials, it is typical to ignore the spin nature and simply account for a factor of two in final result to include spin degeneracy. Here we start off by explicitly accounting for transport in each spin channel.

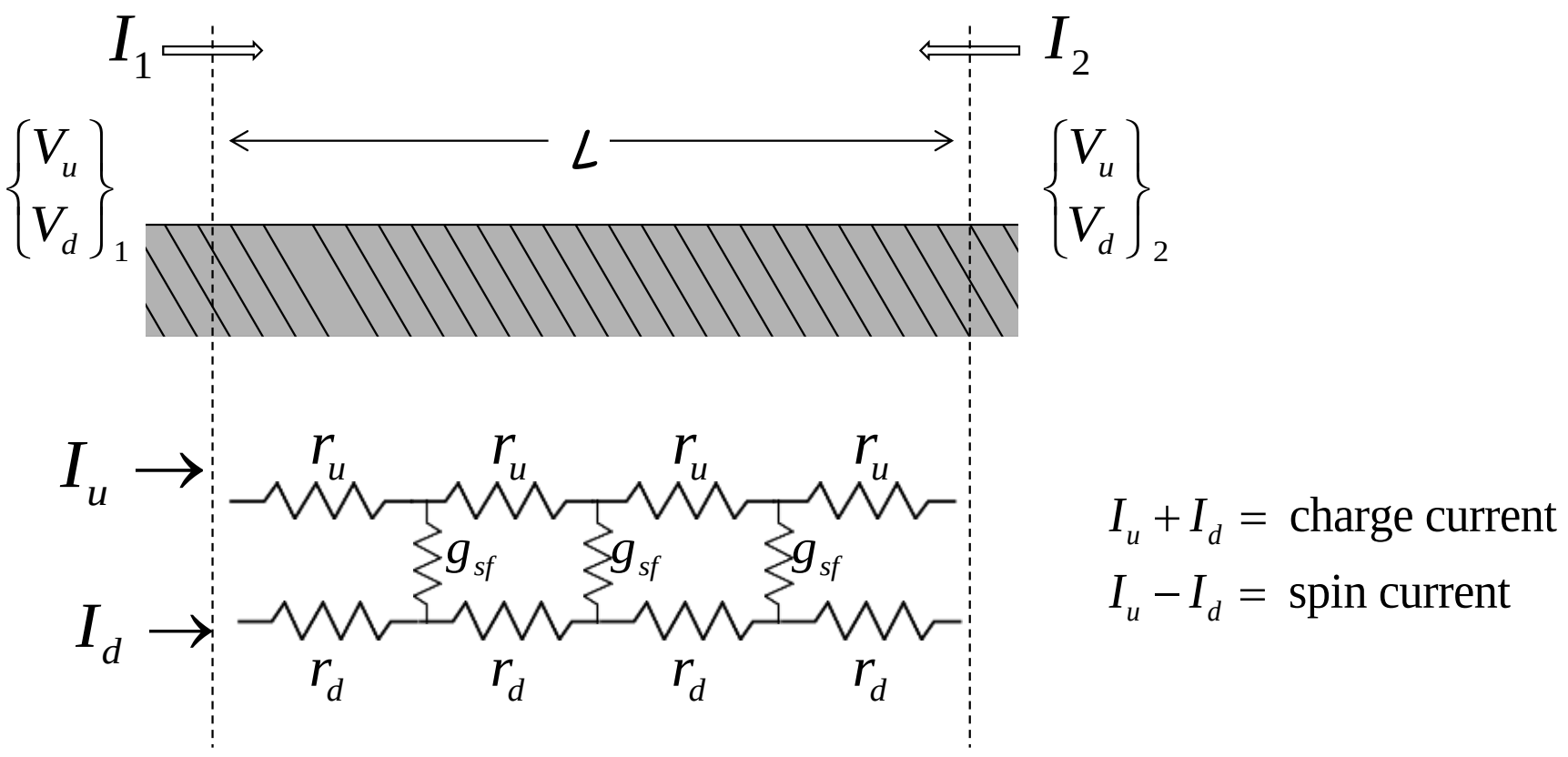

Fig. 8 – Distributed Spin-circuit representation of a spin transport section of length L with boundary conditions $\vec{V}_1$ and $\vec{V}_2$ at the two ends.

For one-dimensional transport the spin-dependent electron current can be written in terms of distributed resistances $r_u$ , $r_d$ and spin-flip conductance $g_{sf}$ [Fig. 8] as

$$r_u I_u = \frac{d(\mu_u / -q)}{dx} = -\frac{dV_u}{dx}, \qquad r_d I_d = \frac{d(\mu_d / -q)}{dx} = -\frac{dV_d}{dx} \tag{2.1a}$$

$$\frac{dI_u}{dx} = -g_{sf}\,(\mu_u - \mu_d)/-q = -\frac{dI_d}{dx} \tag{2.1b}$$

where $\mu_u$, $\mu_d$ are the quasi-Fermi levels for up and down spin channels; $r_{u(d)}$ are the resistances per unit length for each channel in $\Omega - m^{-1}$ and $g_{sf}$ is the spin flip conductance per unit length in $\Omega^{-1}m^{-1}$. These equations can be concisely rewritten in a matrix form as

$$\frac{d}{dx}\begin{Bmatrix} V_u \\ V_d \end{Bmatrix} = -\begin{pmatrix} r_u & 0 \\ 0 & r_d \end{pmatrix}\begin{Bmatrix} I_u \\ I_d \end{Bmatrix}$$
$$\frac{d}{dx}\begin{Bmatrix} I_u \\ I_d \end{Bmatrix} = -g_{sf}\begin{pmatrix} 1 & -1 \\ -1 & 1 \end{pmatrix}\begin{Bmatrix} V_u \\ V_d \end{Bmatrix} \tag{2.2}$$

*Decoupling charge and spin quantities:* When modeling circuits involving the interactions of many magnets that are not necessarily collinear to each other, this representation in the up-down basis has now to be extended to each of the three spatial dimensions, i.e., an up-down representation for each of the 'x', 'y' and 'z' co-ordinates. This makes the situation quite complicated! Modeling such non-collinear systems is greatly facilitated if we can separately consider the charge and spin quantities by using a transformation such as

$$\begin{Bmatrix} V_c \\ V_s \end{Bmatrix} = \frac{1}{2}\begin{pmatrix} 1 & 1 \\ 1 & -1 \end{pmatrix}\begin{Bmatrix} V_u \\ V_d \end{Bmatrix} \text{ and } \begin{Bmatrix} I_c \\ I_s \end{Bmatrix} = \begin{pmatrix} 1 & 1 \\ 1 & -1 \end{pmatrix}\begin{Bmatrix} I_u \\ I_d \end{Bmatrix}$$

where the subscripts 'c' and 's' refer to charge and spin respectively. This basis transformation allows us to rewrite Eq. (2.2) as

$$\frac{d}{dx}\begin{Bmatrix} V_c \\ V_s \end{Bmatrix} = -\frac{1}{4}\begin{pmatrix} r_+ & -r_- \\ -r_- & r_+ \end{pmatrix}\begin{Bmatrix} I_c \\ I_s \end{Bmatrix} \tag{2.3a}$$

$$\frac{d}{dx}\begin{Bmatrix} I_c \\ I_s \end{Bmatrix} = \begin{pmatrix} 0 & 0 \\ 0 & -4g_{sf} \end{pmatrix}\begin{Bmatrix} V_c \\ V_s \end{Bmatrix} \tag{2.3b}$$

where $r_- = r_d - r_u$ ( = 0 for non-magnetic materials, which effectively decouples the charge and spin ) and $r_+ = r_d + r_u$. The advantage of using charge and spin components is that the voltages and currents are conveniently extended to four-component quantities by resolving the spin component into three spatial directions. This becomes trivial in non-magnetic materials noting that there is no distinction between x, y and z components.

*Spin Diffusion equations in the charge-spin basis*: A simple differentiation of Eq. (2.3a) and insubstitution with Eq. 2.3(b) lead to the standard spin diffusion equation now given in the charge-spin basis by:

$$\frac{d^2}{dx^2}\begin{Bmatrix} V_c \\ V_s \end{Bmatrix} = \begin{pmatrix} 0 & -r_- g_{sf} \\ 0 & \lambda_{sf}^{-2} \end{pmatrix}\begin{Bmatrix} V_c \\ V_s \end{Bmatrix} \tag{2.4}$$

where $\lambda_{sf}$ is the spin diffusion length of the channel material given by the relation $\lambda_{sf}^2 = 1/\left(r_+ g_{sf}\right)$.

## 2.2 Lumped Spin Circuit model

The lumped spin circuit model is derived as an analytical solution of the spin diffusion equations [Eqs. 2.3, 2.4)] for any section of length $L$ [Fig. 8] characterized by a resistivity $\rho^{-1} = r_u^{-1} + r_d^{-1}$ , spin diffusion length $\lambda_{sf}$ and with voltages $\vec{V}_1$ and $\vec{V}_2$ across its ends. The solution of Eq. (2.4) can be substituted in Eq. 2.3(a) so that charge and spin currents ($I_c$ and $I_s$) flowing into the section at either end can be related to the voltage drop across its ends ($\Delta V_c$ corresponding to charge and $\Delta V_s$ corresponding to spin) as

$$\begin{Bmatrix} I_c \\ I_s \end{Bmatrix}_1 = \left[G^{se}\right]_{2\times2} \begin{Bmatrix} \Delta V_c \\ \Delta V_s \end{Bmatrix} + \left[G^{sh}\right]_{2\times2} \begin{Bmatrix} 0 \\ V_{s1} \end{Bmatrix} \tag{2.5}$$

The stepwise procedure leading from Eq. (2.4) to Eq. (2.5) is listed in Appendix A and summarized in [Fig. 9] Here we proceed directly to the result, i.e. the structure of the lumped 2x2 series and shunt conductance matrices and how they relate to each of the different regions, namely the non-magnetic channel, the ferromagnet and the interface regions.

Spin Circuit representation:
Evolution from Distributed "Transmission Line" to Lumped element solution

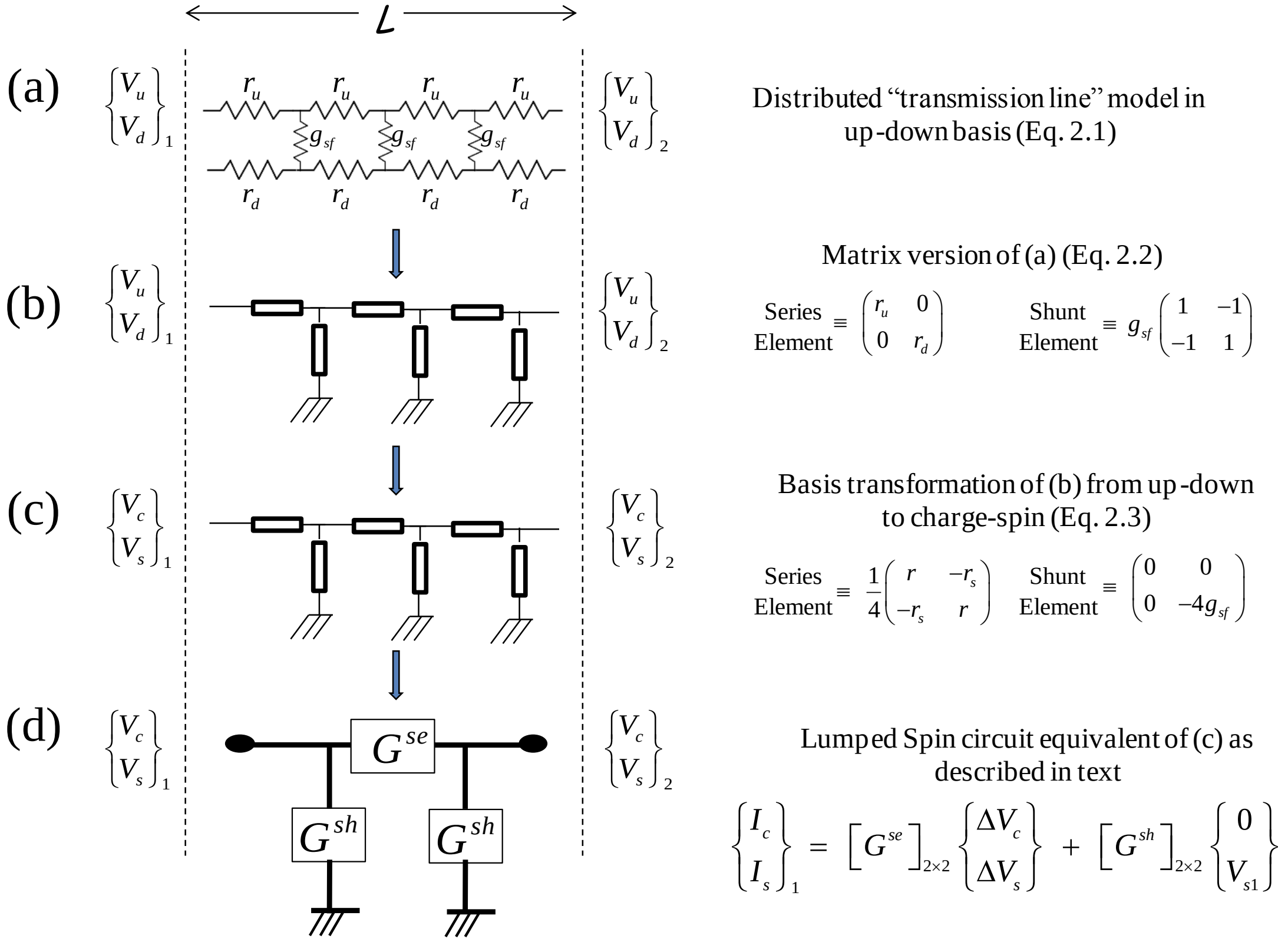

Fig. 9– (a) The distributed network representation of the entire section can be lumped into (d) a pi-network of conductances. Additionally, there is a basis transformation from the standard up-down to a charge-spin basis as explained earlier.

**(A) Non-magnetic Channel**

A non-magnetic material is characterized by an equal number of conducting modes at the Fermi level [Fig. 10] for both the up-spins and down-spins.

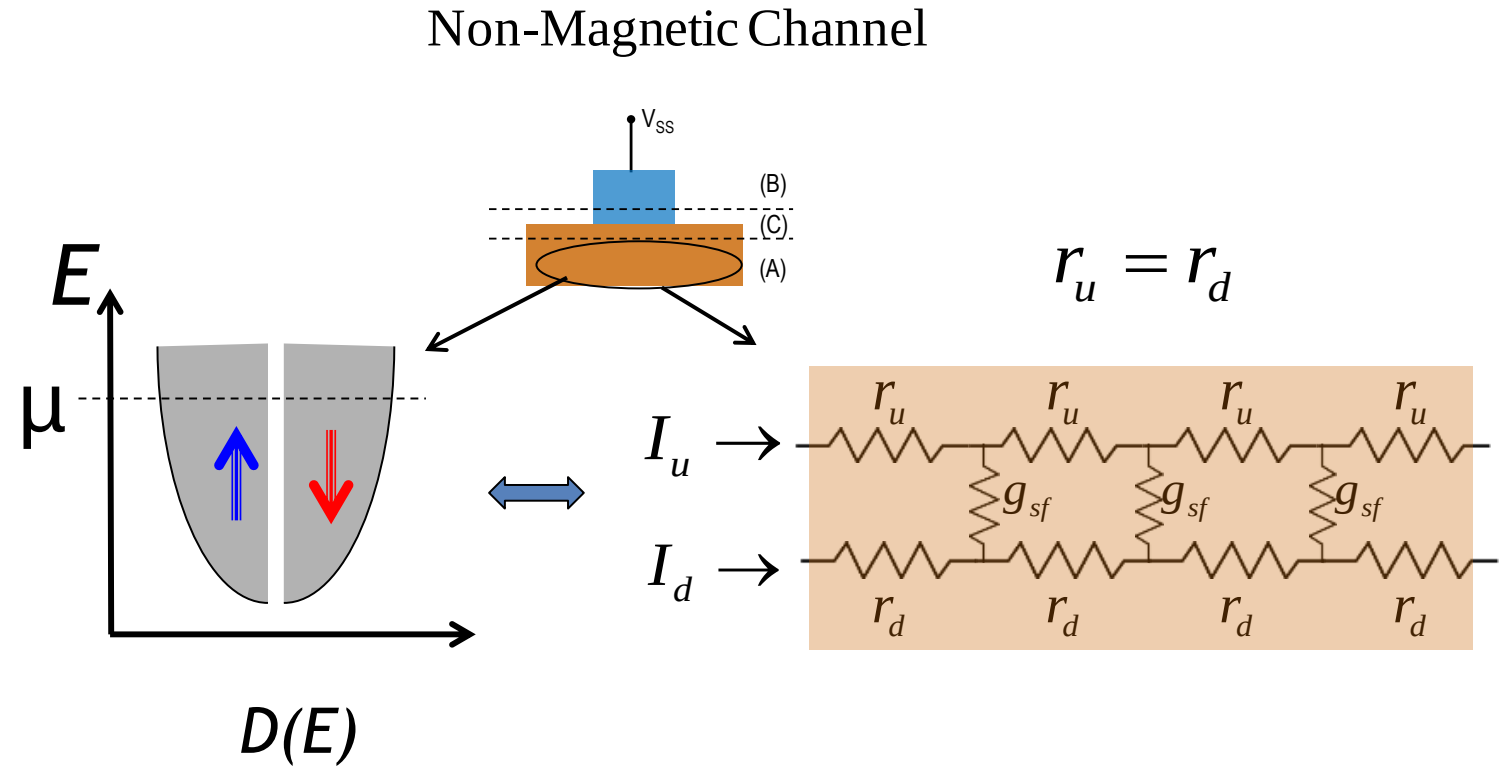

Fig. 10 – Distributed spin circuit representation for a non-magnetic material characterized by the same density of states at the Fermi level for both up and down spins.

Consequently the equivalent circuit model has $r_u = r_d$, i.e. $r_- = 0$.With this simplification in place Eq. (2.4) reduces to :

$$\frac{d^2}{dx^2}\begin{Bmatrix} V_c \\ V_s \end{Bmatrix} = \begin{pmatrix} 0 & 0 \\ 0 & \lambda_{sf}^{-2} \end{pmatrix}\begin{Bmatrix} V_c \\ V_s \end{Bmatrix}$$

Correspondingly, the series and shunt conductance matrices are given by

$$G_N^{se} = \frac{1}{\rho L}\begin{pmatrix} 1 & 0 \\ 0 & \left(\frac{L}{\lambda_{sf}}\right) cosech\left(\frac{L}{\lambda_{sf}}\right) \end{pmatrix}; \quad G_N^{sh} = \frac{1}{\rho L}\begin{pmatrix} 0 & 0 \\ 0 & \left(\frac{L}{\lambda_{sf}}\right) tanh\left(\frac{L}{2\lambda_{sf}}\right) \end{pmatrix};$$

*Series conductance matrix* $\left(G^{se}\right)$**:** Clearly the charge and spin quantities are decoupled due to the absence of any off-diagonal elements in the matrices. The upper diagonal element simply relates the charge voltage drop to the charge current flow by the usual charge conductance as one would expect with Ohm's law. The lower diagonal element relates the spin current flowing through the section to the spin voltage drop across its ends.

*Shunt conductance matrix* $\left(G^{sh}\right)$**:** This matrix has only one element located on the lower diagonal, which is purely spin information. This term is a representation of the spin-flip conductance of a non-magnetic channel. When this matrix is shown to be electrically grounded at one end in the pictorial representation [Fig. 9(d)], we would like to mention that this is not a real ground but rather a *virtual 'spin' ground*. The significance of this matrix is that it captures the spin current dissipation or generation in the structure and is equivalent to the spin-flip conductance $g_{sf}$ in the up-down basis. Unlike charge current, spin current is not conserved and will vary as it keeps encountering spin randomizing events while flowing through any material. This non-conservative nature of the spin current is captured by what flows through the shunt conductance connected to a virtual ground, while the spin polarized current flowing through such a material is captured by the lower diagonal element of the series conductance matrix.

**(B) Ferro-magnet**

A magnet is characterized by an unequal distribution of up-spin and down-spin density of states at the Fermi level [Fig. 11]. This reflects as different conductivities [32] for the up and down spin channels resulting in an unequal flow of up-spin and down-spin currents within the magnet, i.e., a net spin current. Therefore, a magnet possesses the property that a charge current flowing through it is naturally spin polarized.

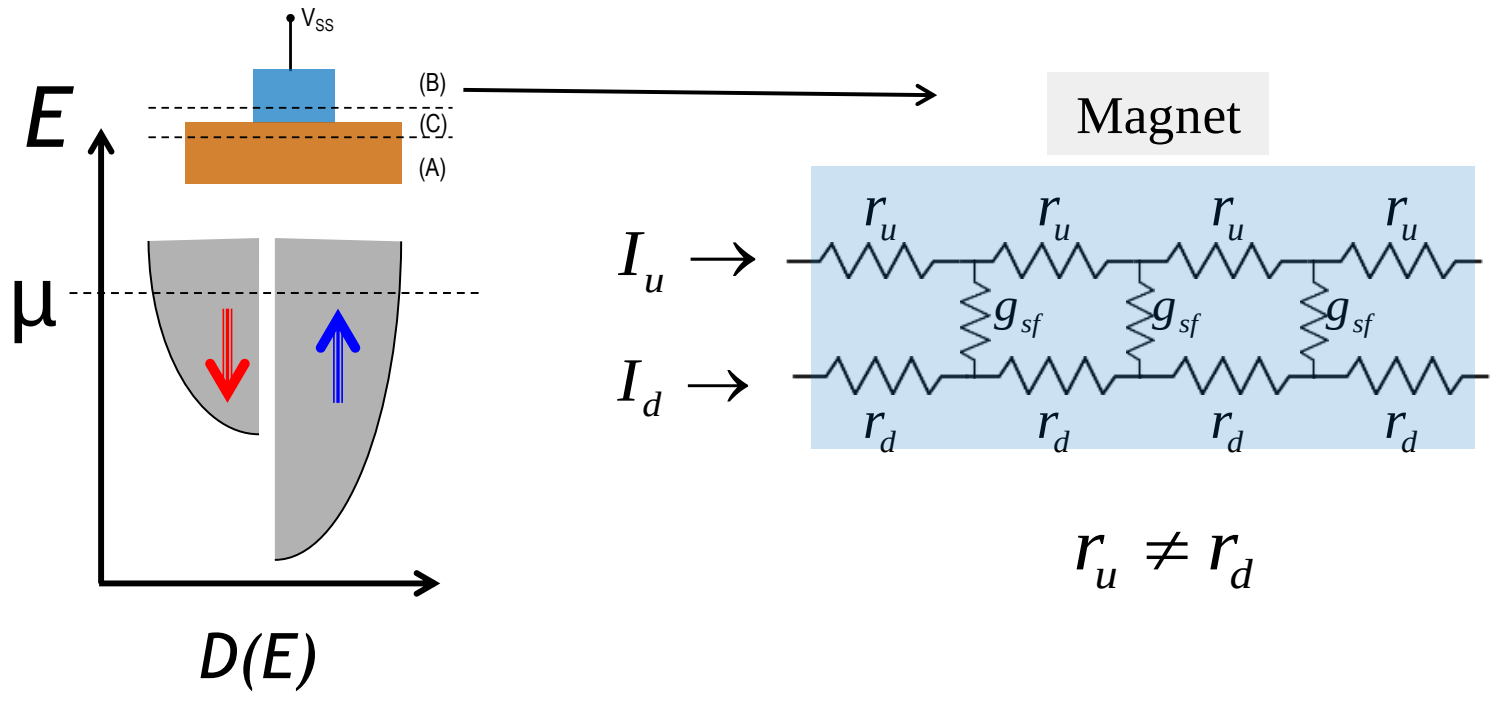

Fig. 11 – Distributed spin circuit representation for a ferro-magnetic material characterized by unequal density of states of up and down spins at the Fermi level.

The ability of the magnet to polarize a charge current is quantified by the parameter '*p*' (spin polarization), defined as $p = |r_u - r_d| / (r_u + r_d)$. Upon solving Eq. (2.4) for the case where $r_u \neq r_d$, this parameter naturally appears (Appendix A) in the final lumped π-network of conductances and couples the charge and spin quantities in the conductance matrix as shown below.

$$G_F^{se} = \frac{1}{\rho L}\begin{pmatrix} 1 & p \\ p & p^2 + \alpha \, cosech\left(\frac{L}{\lambda_{sf}}\right) \end{pmatrix}; \quad G_F^{sh} = \frac{1}{\rho L}\begin{pmatrix} 0 & 0 \\ 0 & \alpha \, tanh\left(\frac{L}{2\lambda_{sf}}\right) \end{pmatrix}; \text{ where } \alpha = \left(1 - p^2\right)\left(\frac{L}{\lambda_{sf}}\right)$$

If these circuit elements are used in describing the flow of current injected into a ferromagnet, then in addition to a charge current flow there is a spin current generated due to this off-diagonal element. On the other hand, for a non-magnetic section, $p = 0$, implying that a charge current flowing through it will not generate a spin current by itself.

**(C) Magnet/Channel Interface region**

The interface between the magnet and the channel can play quite an important role in contributing to the spin polarization of the injected electrons. The first experiments on spin injection dealt with all – metal structures having nearly 'ohmic' interfaces. Later when semiconducting channels were incorporated it was observed that the spin polarization was considerably reduced due to *resistivity mismatch* at the interface. In fact one of the major breakthroughs in experiments is considered to be the introduction of tunnel barriers [33-35] in the interface region. Although it has been conceived that the bandstructure properties of the tunnel barrier would suppress one type of spin carrier (MgO) [36, 37], it is now recognized that even an interface which cannot distinguish between the two types of spins can still help enormously by alleviating the resistivity mismatch (see discussions in[38-40]) between the magnet and the channel.

The mismatch problem can be understood in simple terms keeping the following picture in mind [Fig. 12]. In the case of a magnet injecting into a semiconducting channel through an ohmic interface, it is seen that the number of conducting modes in the magnet is several orders larger than the conducting modes in the channel. Consequently, despite high spin polarization within the magnet, during the process of injection all the modes within the channel are filled, thereby, resulting in zero spin polarization within the channel.

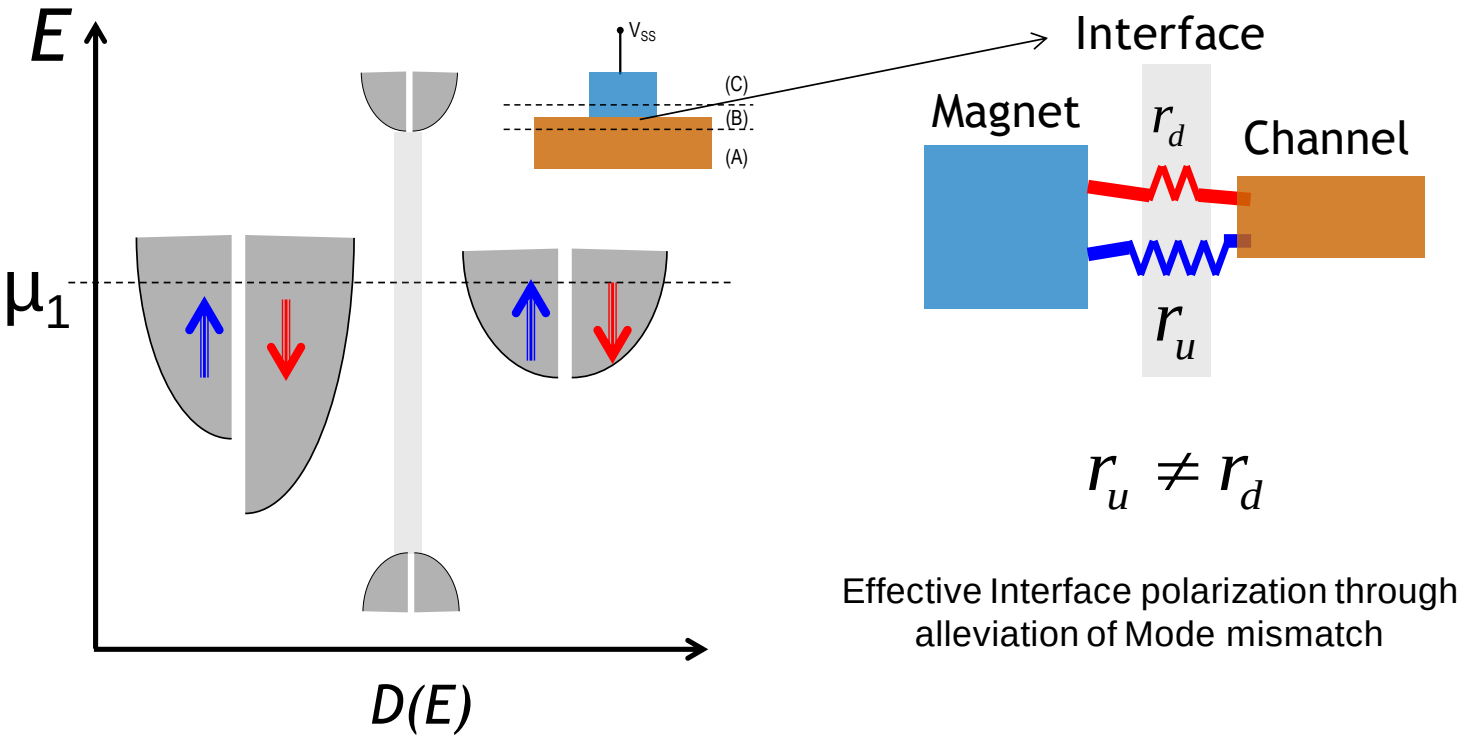

Fig. 12 – Transport through a tunnel-barrier interface between a ferromagnet and channel

A tunnel barrier alleviates this problem by suppressing the overall transmission from the magnet into the channel to such an extent that there is a noticeable difference between the number of up-spin and down-spin carriers that make it through the interface into the channel. Therefore, the interface can be modeled as a region having unequal up and down spin resistances.

Since a circuit model for an ohmic interface does not require any special matrix elements, we will discuss just the modeling of the tunnel barrier. The lumped spin circuit model for the tunnel barrier interface is the same as the one obtained in the case of a magnet. However, it is important to keep in mind that the mechanism of '$p$' is different in these two cases. In general, tunnel barriers at the interface are very thin and it is useful to reduce the conductance matrices with the approximation that $L << \lambda_{sf}$ . This results in the following conductance matrices:

$$G_T^{se} = g_T \begin{pmatrix} 1 & p_T \\ p_T & 1 \end{pmatrix}; \quad G_T^{sh} = [0];$$

where $g_T$ is the conductance of the tunnel barrier and $P_T$ is its effective spin polarization. This use of a constant spin polarization in the conductance matrix for a tunnel barrier region is valid only in the low-bias regime. When injecting across a tunnel barrier at high bias, the bandstructure effects result in a different polarization corresponding to different conducting modes. In this case the effective polarization is determined by integrating over all conducting modes [41].

### 2.3 Example: Spin circuit analysis of non-local spin valve structures

Non-local spin valves [Fig. 13] are among the most popular class of devices used for analyzing spin transport in lateral structures. In this structure a charge current is run through one of the ferromagnets (injector) to a ground terminal. The charge current gets spin polarized by the injecting magnet and this spin current then diffuses towards a detector magnet kept outside the path of the charge current. The presence of the spin current in the channel perturbs the quasi-fermi levels beneath the detector ferromagnet and causes a charge voltage to develop on the detector. The ratio of the measured voltage to the injected current is called non-local resistance ($R_{NL}$). When $R_{NL}$ is a positive number it indicates that the two magnets are parallel to each other and a negative value means that the magnets are anti-parallel.

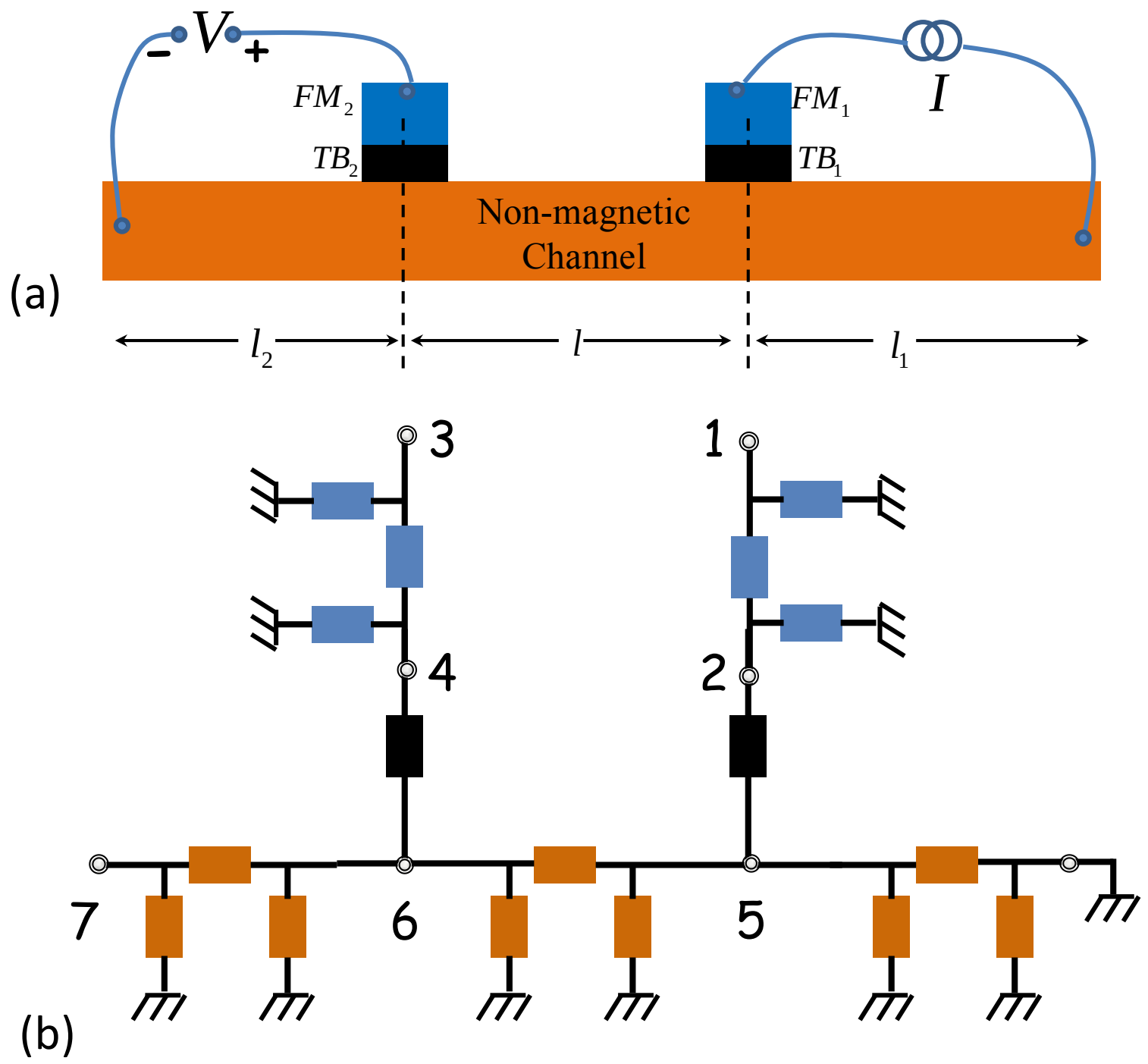

Fig. 13 – Spin circuit representation of a non-local spin valve structure. (a) The non-local spin valve structure can be decomposed into multiple elements such as the injector and detector ferromagnets ($FM_1$ and $FM_2$), tunnel barriers and channel regions. (b) Each element can now be represented by a $\pi$-network of conductances. A simple nodal analysis following current conservation laws can determine the values of currents and voltages at each node.

As an example of lumped spin-circuit analysis we will show how it can be applied to calculate $R_{NL}$ both numerically and analytically. The first step is to break up the spin valve structure into subsections such as shown in [Fig. 13(a)]. The channel portion is decomposed into three parts: a section between the two magnets of length '$l$' and one overhanging region beyond each magnet of length '$l_1$' and '$l_2$' respectively. Similarly, there are subsections corresponding to each magnet and the tunnel barrier interface. Each of these subsections can now be described by an equivalent conductance as shown in [Fig.13 (b)].(Note that the tunnel barrier does not have shunt elements since it is assumed to be much thinner than its spin diffusion length)

Once the equivalent lumped circuit representation has been obtained, the non-local resistance of this structure is given by the ratio of $\left(V_{3c}-V_{7c}\right)/I_{1c}$ , which is the ratio of the charge voltage measured between nodes 3 and 7 to the charge current entering node 1 from the external current source. It is important to note that the circuit representation involving 'virtual' ground terminals allows us to apply the usual current conservation laws for charge currents to spin currents as well. One can represent the total current at any node '$i'$ in the structure as:

$$\sum_j \vec{I}_{ij} = 0 \tag{2.6}$$

where the current $\vec{I}$ is a 2-component vector having a charge and a spin component.
We already noted that the current flowing into any section from a node 'i' is given by Eq. (2.5) as

$$\vec{I}_{ij} = G^{se}\left(\vec{V}_i - \vec{V}_j\right) + G^{sh}\vec{V}_i \qquad \text{(same as Eq. 2.5)}$$

Combining Eqs. (2.5) and (2.6), we can set up a conductance matrix relating all the nodal currents to the nodal voltages in the circuit as:

$$\begin{Bmatrix} \vec{I}_1 \\ \vdots \\ \vec{I}_7 \end{Bmatrix} = \left[G\right]_{total} * \begin{Bmatrix} \vec{V}_1 \\ \vdots \\ \vec{V}_7 \end{Bmatrix} \tag{2.7}$$

Each of these currents and voltages is a 2-component vector since they contain a charge and a spin component. The explicit form of $G_{total}$ is shown below.

$$\begin{pmatrix} G_{F1}+G_{F1} & -G_{F1} & 0 & 0 & 0 & 0 & 0 \\ -G_{F1} & G_{0F1}+G_{F1}+G_{T1} & 0 & 0 & -G_{T1} & 0 & 0 \\ 0 & 0 & G_{0F2}+G_{F2} & -G_{F2} & 0 & 0 & 0 \\ 0 & 0 & -G_{F2} & G_{0F2}+G_{F2}+G_{T2} & 0 & -G_{T2} & 0 \\ 0 & -G_{T1} & 0 & 0 & G_{N1}+G_{0N1}+G_{N2}+G_{0N2}+G_{T1} & -G_{N2} & 0 \\ 0 & 0 & 0 & -G_{T2} & -G_{N2} & G_{N3}+G_{0N3}+G_{N2}+G_{0N2}+G_{T2} & -G_{N3} \\ 0 & 0 & 0 & 0 & 0 & -G_{N3} & G_{N3}+G_{0N3} \end{pmatrix}$$

The subscripts *F*, *T* and *N* correspond to the ferromagnet, tunnel-barrier and non-magnetic regions respectively, while the *G* and $G_0$ correspond to the series $\left(G^{se}\right)$ and shunt $\left(G^{sh}\right)$conductance matrices respectively with reduced subscripts for convenience. The inputs to each of these conductance matrices are the material parameters (resistivity and polarization) and physical dimensions (area, length etc) of the particular section that they describe.

It is relatively straightforward to solve this conductance matrix analytically to determine the ratio $\left(V_{3c}-V_{7c}\right)/I_{1c}$ , though one should keep in mind that each element of is a 2x2 matrix in itself. <u>*Appendix B* provides the details of the analytical derivation of non-local resistance from the above conductance matrix</u>, resulting in the following expression:

$$R_{NL} = \pm 2R_{SN}e^{-L/\lambda_N} \frac{\left(\dfrac{P_{T1}\dfrac{R_{T1}}{R_{SN}}}{1-P_{T1}^2}+\dfrac{P_{f1}\dfrac{R_{F1}}{R_{SN}}}{1-P_{f1}^2}\right)\times\left(\dfrac{P_{T2}\dfrac{R_{T2}}{R_{SN}}}{1-P_{T2}^2}+\dfrac{P_{f2}\dfrac{R_{F2}}{R_{SN}}}{1-P_{f2}^2}\right)}{\left(1+\dfrac{2\dfrac{R_{T1}}{R_{SN}}}{1-P_{T1}^2}+\dfrac{2\dfrac{R_{F1}}{R_{SN}}}{1-P_{f1}^2}\right)\times\left(1+\dfrac{2\dfrac{R_{T2}}{R_{N}}}{1-P_{T2}^2}+\dfrac{2\dfrac{R_{F2}}{R_{SN}}}{1-P_{f2}^2}\right)-e^{-2L/\lambda_N}} \quad (2.8)$$

In this expression, the '$P$'s refer to the spin polarization of the magnet and effective polarization of the tunnel barrier interfaces, while the '$R$'s refer to the resistance of the channel and magnet over one spin diffusion length. The same expression was first derived in Ref. [26] using the spin diffusion equations and is widely cited by experimentalists dealing with these devices.

In practice a quick way to obtain the non-local resistance is to write a simple code to solve this conductance matrix and we have provided a Matlab script in Appendix. C that implements the same procedure detailed above. The simulated result is shown in [Fig. 14] superimposed with the analytical solution from Eq. 2.8.

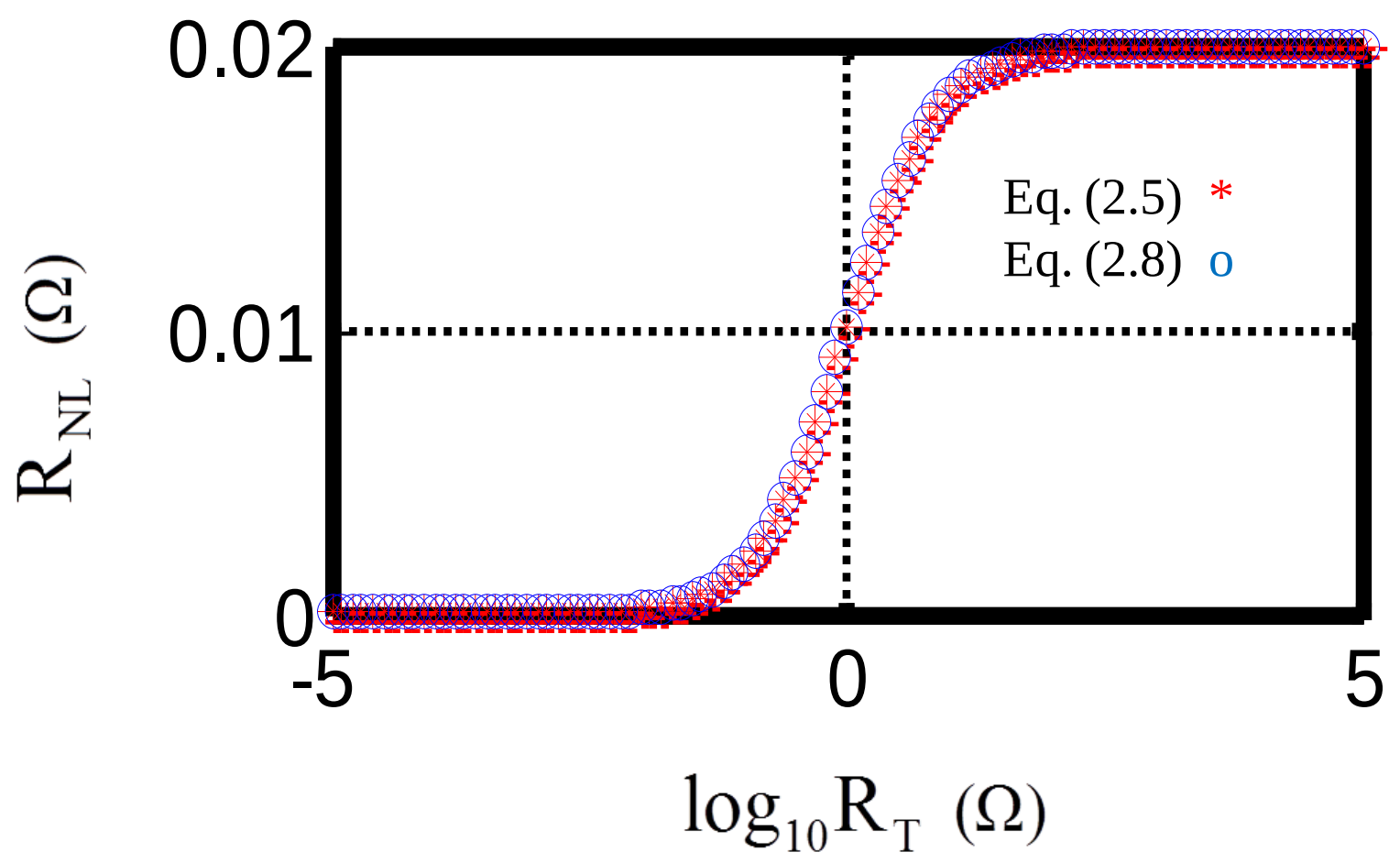

Fig. 14 – Spin circuit solution of the non-local spin valve structure: As an example we plot the dependence of the non-local resistance as a function of the interface resistance assuming all other parameters to be the same. The spin circuit simulation (code provided in appendix) leads to exactly the same result as the analytical expression (Eq. 2.8) derived by Takahashi et al [26].

## 2.4 4 – component Spin circuit representation for non-collinear magnet structures

The preceding example and analysis dealt with the application of lumped spin circuits to a class of devices that involve collinear magnets, i.e., magnets being parallel or anti-parallel to each other. Consequently, the conductance matrices were 2x2 matrices. In order to model non-collinear magnetic systems (such as the experiment discussed in the section 3.2), we now extend our conductances into 4x4 matrices whereby a 4-component voltage drop is related to a 4-component current by a [4 × 4] conductance matrix as: $[I_c, I_s^z, I_s^x, I_s^y] = G^{se}[\Delta V_c, \Delta V_s^z, \Delta V_s^x, \Delta V_s^y]^T + G^{sh}[0, V_s^z, V_s^x, V_s^y]^T$

### (A) Non-magnetic Channel

The 4-component conductance matrices for a non-magnetic section can be obtained by a simple extension of the two component version and would look as follows (ρ: resistivity, L: length , A: cross-sectional area, $\lambda_{sf}$: spin-flip length):

$$G_N^{se} = \frac{A}{\rho L}\begin{pmatrix} 1 & 0 & 0 & 0 \\ 0 & \left(\frac{L}{\lambda_{sf}}\right) cosech\left(\frac{L}{\lambda_{sf}}\right) & 0 & 0 \\ 0 & 0 & \left(\frac{L}{\lambda_{sf}}\right) cosech\left(\frac{L}{\lambda_{sf}}\right) & 0 \\ 0 & 0 & 0 & \left(\frac{L}{\lambda_{sf}}\right) cosech\left(\frac{L}{\lambda_{sf}}\right) \end{pmatrix}$$

$$G_N^{sh} = \frac{A}{\rho L}\begin{pmatrix} 0 & 0 & 0 & 0 \\ 0 & \left(\frac{L}{\lambda_{sf}}\right) tanh\left(\frac{L}{2\lambda_{sf}}\right) & 0 & 0 \\ 0 & 0 & \left(\frac{L}{\lambda_{sf}}\right) tanh\left(\frac{L}{2\lambda_{sf}}\right) & 0 \\ 0 & 0 & 0 & \left(\frac{L}{\lambda_{sf}}\right) tanh\left(\frac{L}{2\lambda_{sf}}\right) \end{pmatrix} \quad (2.9)$$

As we mentioned earlier, the reason for this simple extension is that a non-magnetic material does not distinguish between the x, y and z components of spin.

### (B) Ferromagnet (bulk) region

In the case of a ferromagnet it is important to note that unlike a non-magnetic material, it certainly distinguishes between the different directions of spin. Any spins that are not collinear to the easy-axis of the magnet get randomized within a few monolayers of entering the magnet [21], since the magnet tries to align them along its easy axis. It is this exchange of angular momentum between the non-collinearly incident spins and the magnet that results in a spin torque. From a modeling perspective this means that as long as the magnet is thicker than $\lambda_{sf}$ of the non-collinear components, we can split the magnet into two components: (i) the interface (described in part C) and (ii) the bulk region.

For the bulk region we can heuristically assume that the series conductance corresponding to the non-collinear components is essentially zero. This is usually valid for thicknesses greater than a few nanometers in such materials. The 4-component lumped conductance matrices for a magnet aligned along the 'z' direction are then given by:

$$G_F^{se} = \frac{A}{\rho L}\begin{pmatrix} 1 & p & 0 & 0 \\ p & p^2 + \alpha\, cosech\left(\frac{L}{\lambda_{sf}}\right) & 0 & 0 \\ 0 & 0 & 0 & 0 \\ 0 & 0 & 0 & 0 \end{pmatrix}_{c\ z\ x\ y}; \quad G_F^{sh} = \begin{pmatrix} 0 & 0 & 0 & 0 \\ 0 & \left(\frac{A}{\rho L}\right)\alpha\, tanh\left(\frac{L}{2\lambda_{sf}}\right) & 0 & 0 \\ 0 & 0 & g'_{sf} & 0 \\ 0 & 0 & 0 & g'_{sf} \end{pmatrix}_{c\ z\ x\ y}$$

$$\alpha = \left(1 - p^2\right)\left(\frac{L}{\lambda_{sf}}\right) \tag{2.10}$$

The upper left 2x2 quadrant remains the same as the 2-component case earlier for both the series and shunt matrices. The 'x' and 'y' components are set up to reflect that any non-collinear component of spin current entering a magnet is dissipated and can be numerically quantified by the current flowing through a large shunt conductance $G_{FM}^{sh}$ connected to a virtual ground. In the shunt conductance matrix we can heuristically set a large spin flip conductance $g'_{sf}$ that can be related to the relaxation time of transverse spins in the magnet. For modeling purposes, this does not matter very much since any action related to the non-collinear part takes place at the interface and does not appear in the bulk region. A more in-depth discussion appears in [22]. Of course, this qualitative argument is valid for a relatively thick magnet. In the case of a very thin magnet, the situation might vary [42] since some of the non-collinear components of spin currents may successfully traverse the magnet and we may need a more detailed treatment for obtaining an accurate model.

**(C) Channel-Magnet Interface region**

In order to describe transport across the interface from a non-magnetic material into the first few monolayers of the magnet, we use a slightly different approach, pioneered by Brataas et al. [21]. The components of the interface conductance matrix can be derived from scattering theory to describe transport between a plane inside the channel and a plane inside the magnet. Following the work in [21] we will show in Appendix B that for a magnet pointing in the 'z' direction, scattering theory leads to conductance matrices defined below in Eq. 2.11.

$$G_{Int}^{se} = \frac{q^2}{h} M \begin{pmatrix} 1 & P & 0 & 0 \\ P & 1 & 0 & 0 \\ 0 & 0 & 0 & 0 \\ 0 & 0 & 0 & 0 \end{pmatrix} \begin{vmatrix} c \\ z \\ x \\ y \end{vmatrix}; \quad G_{Int}^{sh} = \frac{q^2}{h} M \begin{pmatrix} 0 & 0 & 0 & 0 \\ 0 & 0 & 0 & 0 \\ 0 & 0 & a & b \\ 0 & 0 & -b & a \end{pmatrix} \begin{vmatrix} c \\ z \\ x \\ y \end{vmatrix} \tag{2.11}$$

$a \approx 1$ and $b \approx 0$ for ohmic interfaces

where M is the number of conducting modes at the interface given by $M = k_f^2 A / 4\pi$ and $k_f$ is the wave-vector in the channel. A cursory observation of Eq. (2.11) shows that the upper left quadrant is equivalent to the one obtained for a bulk magnetic section (Eq. 2.10) with L << $\lambda_{sf}$. The shunt conductance matrix for the interface contains components corresponding to the 'x' and 'y' direction and quantifies the spin torque acting on the magnet. Since we do not wish to diverge into the theory of spin torque, we will merely point out here that the diagonal elements refer to the 'Slonczewski' term while the off-diagonal elements refer to the 'field-like' term for spin torque. Typically, for experiments involving ohmic interfaces between the magnet and the channel, the field like term is believed to be negligible [23] and, if the interface is a very clean one, we can assume that the interface conductance is close to the ballistic limit. For treatment of more complex interfaces such as a tunnel barrier with ferromagnetic insulator, we refer the reader to the detailed work of [23] and references therein.

**Basis transformation for an arbitrary direction $(\hat{m})$:**

When modeling circuits involving non-collinear magnets there is one additional step we have to account for. We mentioned that Eqs. (2.10 and 2.11) were derived for a magnet assuming that its easy axis lies in the 'z' direction. As long as we are dealing with just one magnet, it does not really matter what we consider as the 'z' axis. However, with multiple non-collinear magnets, where each one has its own 'z' direction (corresponding to the easy-axis), it has to be ensured that the conductance matrices for all the magnets are written in a single uniform basis. In practice, a simple way to accomplish this is to initially construct the conductance matrices for each magnet according to Eqs. 2.10 and 2.11, and subsequently perform a unitary rotation operation to reflect the actual directions in which the magnets are pointing. The operation looks as follows

$$\begin{aligned} G_F^{se}(\hat{m}) &= U(\hat{z} \to \hat{m})\; G_F^{se}(\hat{z})\; U^{\dagger}(\hat{z} \to \hat{m}) \\ G_F^{sh}(\hat{m}) &= U(\hat{z} \to \hat{m})\; G_F^{sh}(\hat{z})\; U^{\dagger}(\hat{z} \to \hat{m}) \end{aligned} \qquad (2.12)$$

where the rotation operator $U$ is used to transform the conductance matrix from the $\hat{z}$ direction to $\hat{m}$. The formulation of $U$ is given by the Rodrigues Rotation formula and is derived as follows.

In order to construct a general rotation matrix operator $U(\hat{m}_1 \to \hat{m}_2)$ that rotates a vector along a direction $\hat{m}_1$ into one along a direction $\hat{m}_2$ separated by an angle $\theta$, it is convenient to first define a unit vector $\hat{u}$ that is perpendicular to the plane containing $\hat{m}_1$ and $\hat{m}_2$ such that $\hat{u} = \hat{m}_1 \times \hat{m}_2 / |\hat{m}_1 \times \hat{m}_2|$.

The rotation operator $U$ is then given by

$$U(\hat{m}_1 \to \hat{m}_2) \equiv \begin{pmatrix} 1 & 0 & 0 & 0 \\ 0 & u_z^2 + (1-u_z^2)c & u_x u_z(1-c) - u_y s & u_y u_z(1-c) + u_x s \\ 0 & u_x u_z(1-c) + u_y s & u_x^2 + (1-u_x^2)c & u_x u_y(1-c) - u_z s \\ 0 & u_y u_z(1-c) - u_x s & u_x u_y(1-c) + u_z s & u_y^2 + (1-u_y^2)c \end{pmatrix} \begin{matrix} c \\ z \\ x \\ y \end{matrix}$$

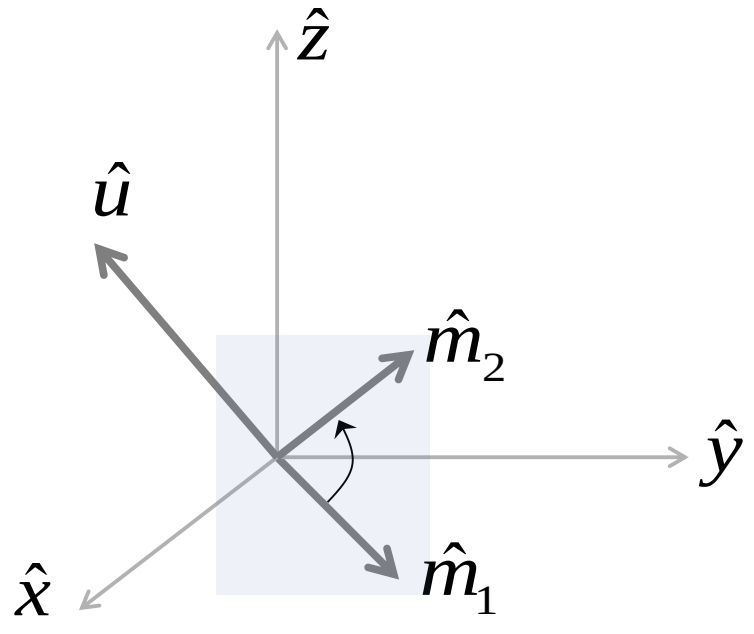

# 3. A coupled spin-transport/ magnetization-dynamics simulator

Once the spin circuit for a multi-magnet network has been set up as described in the previous section, it provides the information to the magnets in the form of spin currents. The circuit now has to be coupled to a magnetization dynamics simulator [Fig. 15], which determines how the magnetization of the magnets responds to input spin information. This part of the model relates to how (spin) information is processed in the nanomagnets and the physical phenomenon responsible for this processing is spin torque switching. The formalism used here is the standard Landau-Lifshitz-Gilbert (LLG) equation with the Slonczewski and field like terms included for spin torque. The magnets are assumed to be monodomain, since we envision nanometer sized magnets in realistic devices. We will briefly describe the LLG block and show how to couple it to the spin-circuit model in this section.

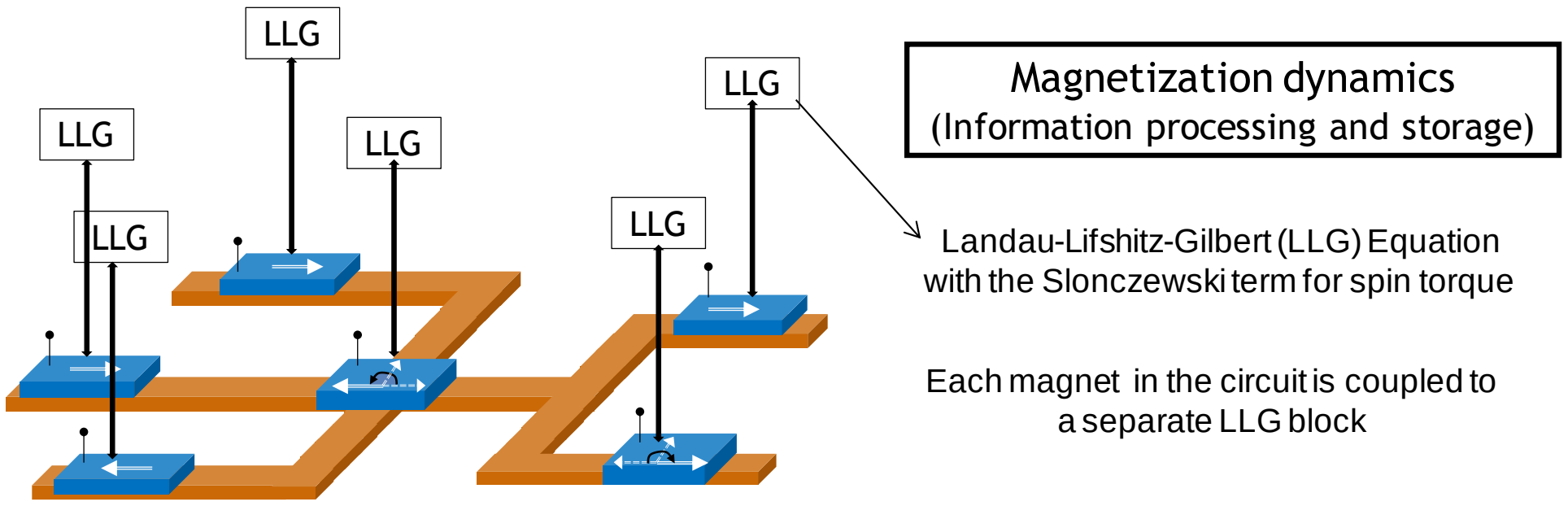

Fig. 15 – LLG solver for the circuit in Fig. 1. In this chapter we will provide a simple code that solves the LLG equation for magnets that are assumed to be monodomain and free of thermal noise effects.

## 3.1 LLG solver for magnetization dynamics

The LLG solver computes the solution to the following dynamical equation, which describes the instantaneous magnetization ($\hat{m}$) in the presence of external perturbation such as magnetic fields or spin currents.

$$(1+\alpha^2)\frac{d\hat{m}}{dt} = -|\gamma|\left(\hat{m}\times\vec{H}\right) - \alpha|\gamma|\left(\hat{m}\times\hat{m}\times\vec{H}\right) + \vec{\tau} + \alpha\left(\hat{m}\times\vec{\tau}\right)$$

$$\text{where } \vec{\tau} = \frac{\hat{m}\times\vec{I}_S\times\hat{m}}{qN_s} \equiv \text{Spin torque} \tag{3.1}$$

The above equation is written in CGS units and the assumption here is that the magnet is mono-domain and can be characterized by a single ($\hat{m}$). The fixed parameters in the equation include the following: $\gamma$ is the gyromagnetic ratio (17.6 MHz / Oersted), $\alpha$ is the Gilbert damping parameter (specific to each magnet and determined from experiment), $q$ is the charge of an electron and $N_s$ is the total number of spins in the nanomagnet given by the relation $N_s = M_s\Omega/\mu_B$ ($M_s$: saturation magnetization, $\Omega$: volume and $\mu_B$: Bohr magneton).

$\vec{H}$ represents the sum of the internal and external fields on the magnet. In the absence of any external fields there are still internal fields present, which are what are responsible for keeping the magnetization pointing along the easy axis. For example, a thin film magnet oriented in the x-z plane with easy axis along ($\hat{z}$) is characterized by $\vec{H} = H_K m_z\hat{z}$ - $H_d m_y\hat{y}$ representing the internal "uniaxial anisotropy" and "out-of-plane demagnetizing" effective fields.

The last two terms relate to spin torque. $\vec{I}_s$ is the spin current provided by the spin-circuit and the definition of $\tau$ clearly implies that only those components of $\vec{I}_s$ perpendicular to $\hat{m}$ contribute towards spin torque. Also this term indicates that the effect of spin torque is greater when $\vec{I}_s$ is increased [Fig. 16] or $N_s$ is reduced, i.e. by making the magnet smaller [14].

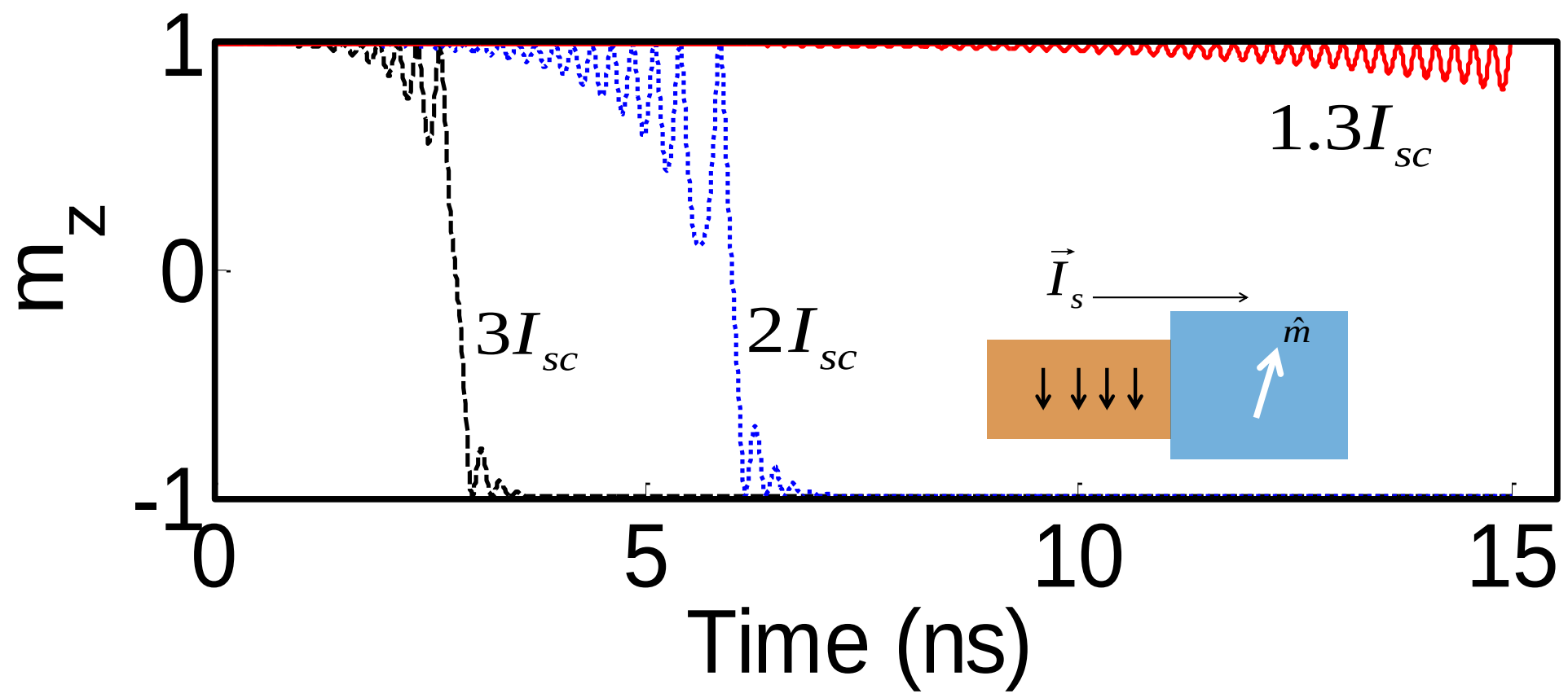

Fig. 16 – LLG solver describing spin torque switching. An increase in the input spin current (denoted in terms of the critical spin current required for switching) reduces the switching time.

In the appendix we provide a code to reproduce [Fig. 16] by utilizing the implicit ODE solver available in MATLAB to solve the LLG equation. The input to the LLG block is an assumed constant spin current in the '–z' direction (easy axis). As an example we consider 3 different magnitudes of the input current expressed in terms of the critical spin current [43] required for spin torque switching about the easy axis of the magnet. The simulation shows that increasing the spin current, above the critical value, reduces the switching time of the magnet along its easy axis.

### 3.2 Coupling spin transport with magnetization dynamics

When coupling the spin-circuit block to the LLG block, the important point to note is that the LLG equation is a dynamical one, while the spin-circuit block provides a steady-state analysis of spin transport.

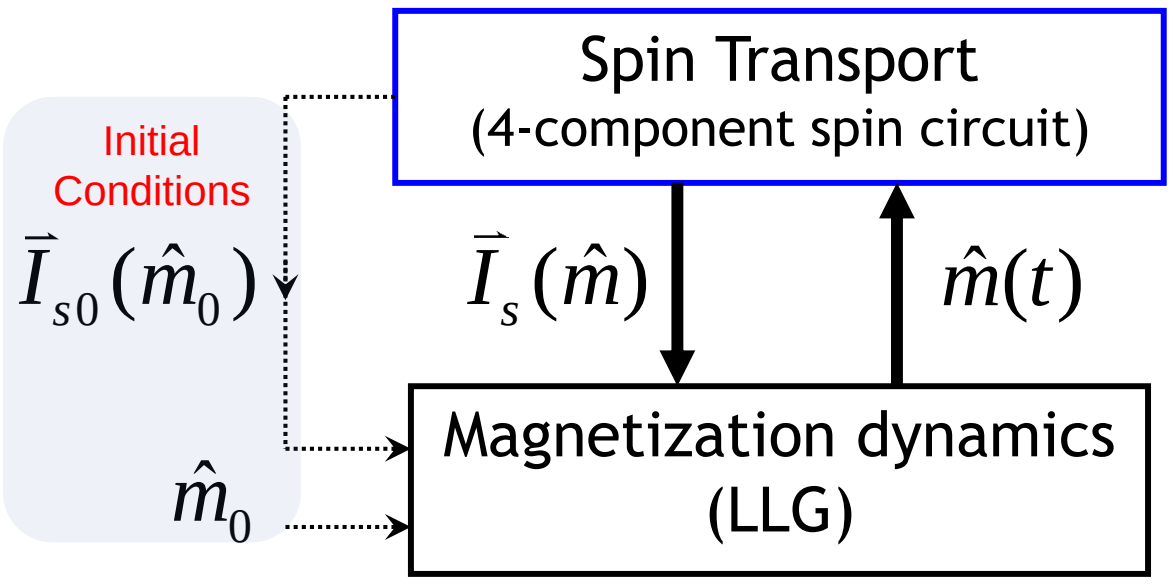

Fig. 17 – Coupled spin-transport/magnetization-dynamics solver. The LLG solver describes magnetization at every time instant m(t) as an input to the spin circuit. The spin circuit uses this value of m to update the spin currents using a steady state analysis.

The way we approach the problem is that for every instant that the magnet is turning, the instantaneous magnetization direction ($\hat{m}$) is supplied to the spin-circuit block [Fig. 17]. This value of $\hat{m}$ is used to update the conductance matrices according to Eq. (2.12). The spin-circuit block then recalculates the spin currents flowing into the magnet and supplies this back to the LLG block, which then proceeds with the magnetization dynamics until the next time step and so forth. This whole process of simultaneously solving the spin-transport and magnetization dynamics continues until the magnet settles to a preferred stable state. This approach is very accurate when dealing with systems where the transit time of spins within the transport channels is much shorter than the switching time of the nanomagnet. For channels of a few 100nm in length, this transport duration can be in pico-seconds whereas present-day magnets switch close to a 100 pico-seconds at best. However, if these times (spin transport and magnet dynamics) become comparable as this technology advances in the future, then one would have to include a dynamical description of spin transport.

**Experiment and Benchmark:** We will now illustrate how the coupled model can be used to benchmark the experiment in Ref. [15]. This structure is physically identical with the non-local spin valve that we analyzed earlier in section 2.3. There is a permalloy magnet injecting spin current into a copper channel. This spin current flows towards the detector permalloy magnet and is measured as a charge voltage. The magnets are connected to the external measurement circuit via gold leads. Unlike the example discussed in section 2.3, the magnets here are not entirely collinear due to the presence of some random fluctuations as is expected in practice. Consequently the non-local spin current exerts a spin torque on the detector magnet. As the current from the external source is ramped up, the detector magnet switches beyond a certain critical current Fig. 18. This switching behavior is modeled by the LLG block, which models the X-axis of the experimental result. The switching is reflected by a change in the sign of the non-local resistance (Y-axis), which is modeled by the spin-circuit block

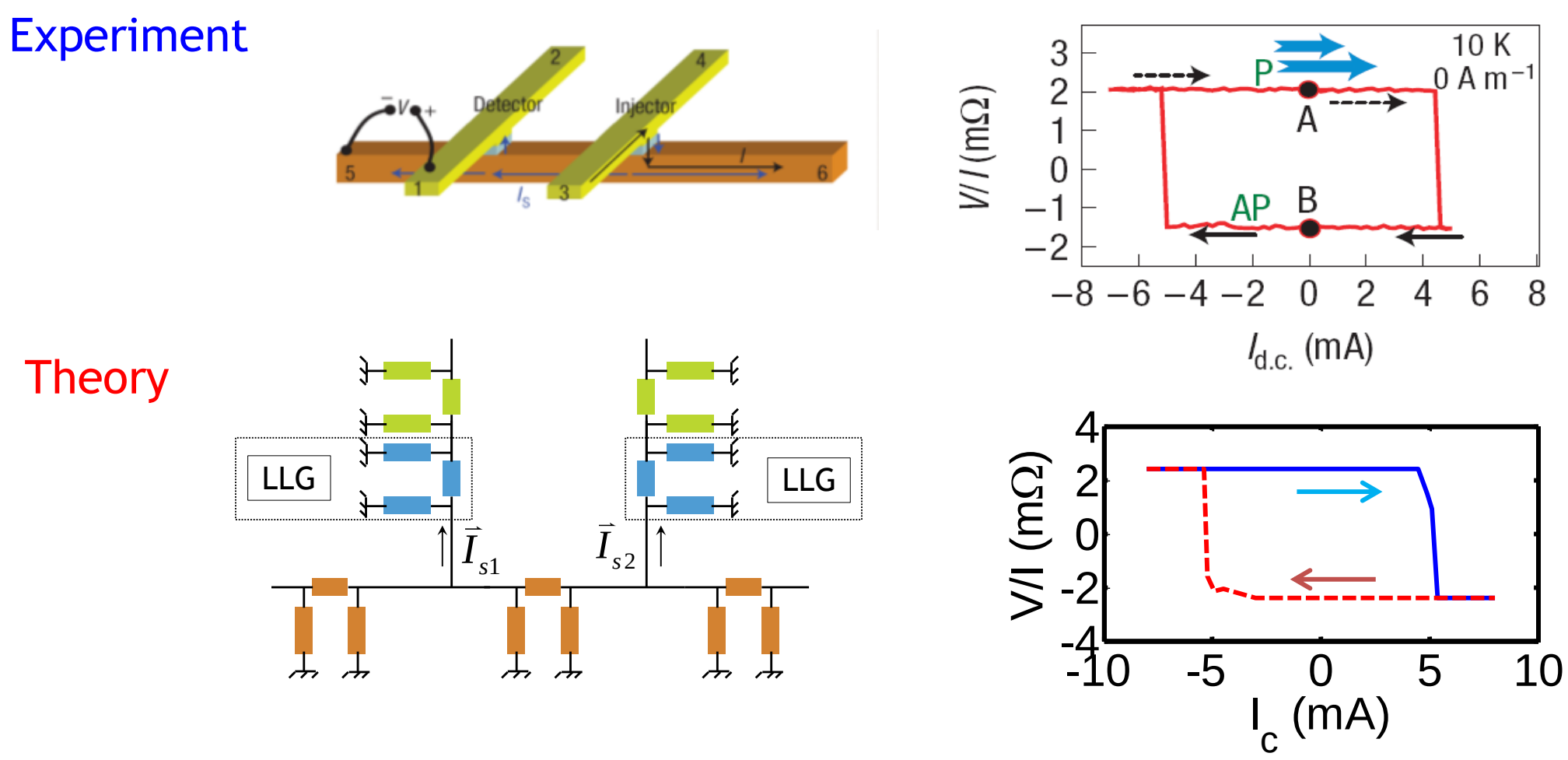

Fig. 18: 4-component spin circuit solution of an experiment (reproduced with permissions from [14]) involving spin-torque switching in a non-local spin valve. The y-axis is the non-local resistance, which requires the same analysis as in example 1. However, the x-axis is indicative of spin-torque switching, which is all about non-collinear spin currents and requires a 4-component spin circuit analysis.

A Matlab code demonstrating this coupled spin-transport/ magnetization-dynamics solution shown in Fig. 18 is included in the Appendix for the reader's reference. For the spin circuit part, the inputs to the conductance matrices are the material properties and physical dimensions of the structure while the polarization of the magnets was adjusted to a value of ~0.5, which is in a reasonable range. The conductance matrix for the magnet shown in the Matlab code is a combination of those for bulk magnet and the interface. Since the thickness of the magnet places it in the diffusive regime, the series conductance is determined by the bulk value. The shunt conductance on the other hand is determined by the interface conductance matrix since all the non-collinear action takes place around the interface.

Assuming that the magnets are initially oriented along $\hat{z}$ with slight deviation due to thermal fluctuation, the spin circuit computes the various currents in the structure. The spin currents $\vec{I}_{s1}$ and $\vec{I}_{s2}$ entering the magnets are fed to the LLG block, which then computes the effect of these currents on the magnetization and returns updated values of magnetization back to the spin circuit block. This process continues till the magnets stabilize to a final state.

### 3.3 Simulating Multi-magnet networks interacting via spin currents

The real advantage of the coupled spin-circuit/LLG model becomes apparent when we have to model networks of interacting magnets and have to describe their magnetizations simultaneously in real time [Fig. 19].

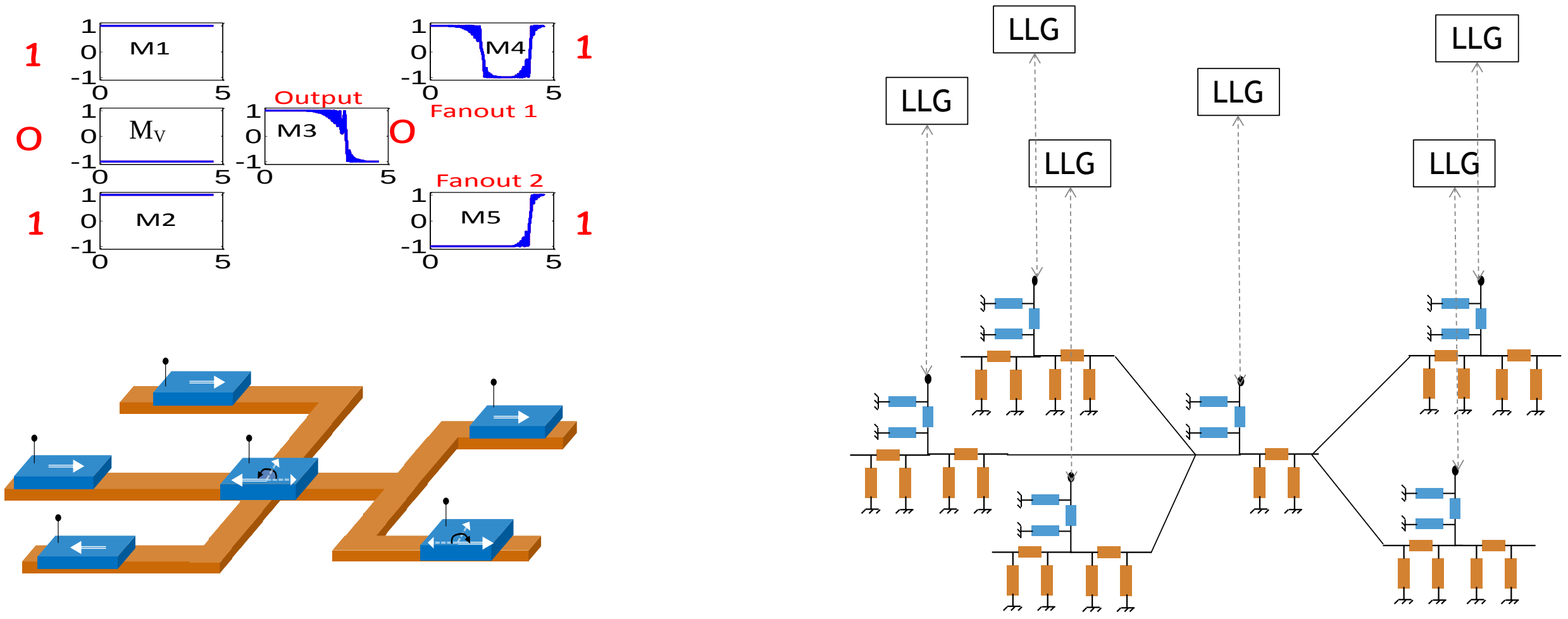

Fig.19 – Coupled spin-transport/magnetization-dynamics simulator for multi-magnet networks. The example shown here is a NAND gate implementation using ASL.

Consider the example of a multi-magnet NAND logic gate implemented using ASL [19]. This NAND architecture is quite different from standard CMOS based implementation as it is based on the majority logic functionality. The spin signals from the inputs add up in an analog fashion and if the resultant is greater than a certain threshold, the output magnet (M3) switches its magnetization. Although M3=NAND(M1,M2), M3 has a third input ($M_V$) which is what creates a majority. For the NAND operation $M_V$ is set to ‘0’ while a NOR operation is achieved if is set to ‘1’.

Simulations in [Fig. 19 ] show the dynamics of all the magnets for a particular choice of inputs (both logic level high (1)), essentially illustrating one row of the NAND gate truth table. The output magnet (M3) inverts the majority of the input and goes to the ‘0’ state. M3, in turn, inverts the fan-out magnets (M4, M5) clearly showing the directed transfer of information from the input to the output to the fan-out stages. How this directionality is achieved and the design parameters that go into achieving basic ASL device level operation is explained in [13, 18]. Here we just wish to illustrate the point that, once we have the basics of the coupled model in place, it becomes a valuable tool for investigating circuit design involving several magnets interacting together.

## 4. Concluding remarks

Information processing through spin-magnet systems is based on two key recent advances namely (1) the demonstration of spin injection into metals and semiconductors from magnetic contacts and (2) the switching of a magnet by the injected spins, which provide mechanisms for reading and writing respectively. However, for the purpose of performing logic based computations, such *Read* and *Write* processes can be combined to implement large scale circuits only if individual W-R units can be designed to exhibit a transistor-like gain and directivity. The ASL concept represents a practical first step in this evolution from physical principles of spin transport/magnetization dynamics to the design of logic devices. In our earlier work we have designed and analyzed such ASL circuits, and this work has been extended towards ASL – based Integrated Circuit simulation frameworks by other groups [44].

*The purpose of this chapter* is to describe in detail the experimentally benchmarked coupled spin circuit/magnetization dynamics formalism that we have developed for analysis of spin-magnet systems. Our modeling approach has broad applicability to analyzing and providing guidelines to existing spin valve [24] and spin torque [45] experiments. It can also be employed for designing other experiments that could be more accessible on a shorter timeframe; such as Ref. [46] that proposes a new class of probabilistic experiments that could be performed with stochastic spin-magnet circuits.

## Acknowledgement

S.S. was supported by the Institute for Nanoelectronics Discovery and Exploration (INDEX), while A.S. and V.D. were supported by the Center for Science of Information (CSoI), an NSF Science and Technology Center.

## A. Derivation of lumped representation of Spin Circuit (Eq. 2.5)

This appendix lists the derivation of the lumped pi-network of conductance matrices given by Eq. (2.5), i.e.,

$$\begin{Bmatrix} I_c \\ I_s \end{Bmatrix}_1 = \left[G^{se}\right]_{2\times 2} \begin{Bmatrix} \Delta V_c \\ \Delta V_s \end{Bmatrix} + \left[G^{sh}\right]_{2\times 2} \begin{Bmatrix} 0 \\ V_{s1} \end{Bmatrix}$$

$$\text{where } G^{se} = \frac{1}{\rho L}\begin{pmatrix} 1 & p \\ p & p^2 + \alpha\, cosech\left(\frac{L}{\lambda_{sf}}\right) \end{pmatrix}; \quad G^{sh} = \frac{1}{\rho L}\begin{pmatrix} 0 & 0 \\ 0 & \alpha\, tanh\left(\frac{L}{2\lambda_{sf}}\right) \end{pmatrix}; \alpha = \left(1 - p^2\right)\left(\frac{L}{\lambda_{sf}}\right) \tag{2.5}$$

which is obtained as an analytical solution to the spin diffusion equations given by Eqs. (2.3 and 2.4). To do so let us first recap the spin diffusion equations given in the matrix form as

$$\frac{d}{dx}\begin{Bmatrix} V_c \\ V_s \end{Bmatrix} = -\frac{1}{4}\begin{pmatrix} r_+ & -r_- \\ -r_- & r_+ \end{pmatrix}\begin{Bmatrix} I_c \\ I_s \end{Bmatrix} \tag{A1}$$

$$\frac{d^2}{dx^2}\begin{Bmatrix} V_c \\ V_s \end{Bmatrix} = \begin{pmatrix} 0 & -r_- g_{sf} \\ 0 & \lambda_{sf}^{-2} \end{pmatrix}\begin{Bmatrix} V_c \\ V_s \end{Bmatrix} \tag{A2}$$

where $\lambda_{sf}^2 = 1/\left(r_+ g_{sf}\right)$. The procedure is quite straightforward. Eq. (A2) can be solved for any section of length L with specified values of $\Delta V_C$ and $\Delta V_s$ across its ends and putting this solution back into Eq. (A1) results in an expression for the charge and spin currents ($I_C$ and $I_s$) at either end in terms of the voltages given by Eq. (2.7). The details of the procedure are given below.

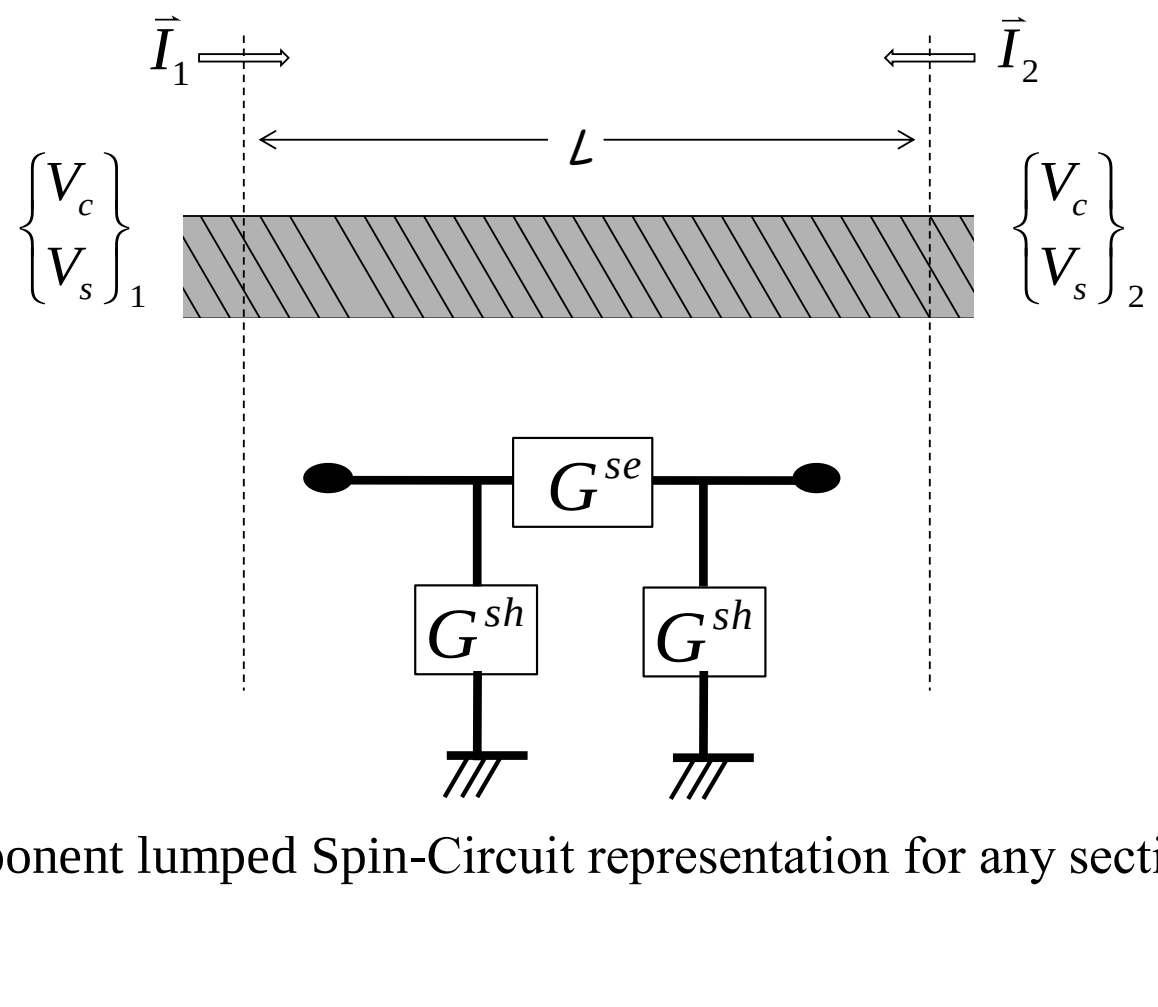

Fig. A.1. 2-component lumped Spin-Circuit representation for any section of length ‘L’.

Step 1: Solving Eq. (A2)

Row 1 of Eq. (A2) determines the charge voltage an any point along the structure

$$\begin{aligned} V_c &= Ax + B(L-x) - r_- g_{sf} \int dx \int dx\, V_s \\ &= Ax + B(L-x) - \left(r_- g_{sf} \lambda_{sf}^2\right) V_s \\ &= Ax + B(L-x) - PV_s \qquad \because r_- g_{sf} \lambda_{sf}^2 = \frac{r_-}{r_+} = P \end{aligned}$$

Here $P$ refers to the effective polarization of the conducting section defined by $P = \frac{g_u - g_d}{g_u + g_d} = \frac{r_d - r_u}{r_d + r_u}$

Solving for the constants $A$ and $B$ by plugging in the boundary conditions $x=0$ *and* $x=L$ respectively, we obtain

$$V_1 = BL - PV_{s1} \quad \text{and} \quad V_2 = AL - PV_{s2}$$

$$\therefore V_c = \left(V_2 + PV_{s2}\right)\frac{x}{L} + \left(V_1 + PV_{s1}\right)\frac{L-x}{L} - PV_s \tag{A3}$$

Similarly, the solution for row 2 of Eq. (A2) gives the expression for the spin voltage as

$$V_s = \frac{V_{s2}\sinh(\frac{x}{\lambda_{sf}}) + V_{s1}\sinh(\frac{L-x}{\lambda_{sf}})}{\sinh(\frac{L}{\lambda_{sf}})} \tag{A4}$$

Step 2: Writing expressions for current from Eq. (A1)

Turning towards Eq. (A1) one can also solve for the currents in the structure as:

$$I_c = \frac{-4\left(r_+ \frac{dV_c}{dx} + r_- \frac{dV_s}{dx}\right)}{r_+^2 - r_-^2} \quad ; \qquad I_s = \frac{-4\left(r_- \frac{dV_c}{dx} + r_+ \frac{dV_s}{dx}\right)}{r_+^2 - r_-^2}$$

These expressions for current can be written in a more concise fashion using the following algebraic simplifications:

$$\frac{4r_+}{r_+^2 - r_-^2} = \frac{r_u + r_d}{r_u r_d} = r_u^{-1} + r_d^{-1} = g_u + g_d = g$$

$$\frac{4r_-}{r_+^2 - r_-^2} = \frac{r_d - r_u}{r_u r_d} = r_u^{-1} - r_d^{-1} = g_u - g_d = Pg$$

This leads to simplified expressions for current as

$$I_c = -g\left(\frac{dV_c}{dx} + P\frac{dV_s}{dx}\right); \qquad I_s = -g\left(P\frac{dV_c}{dx} + \frac{dV_s}{dx}\right) \tag{A5}$$

Step 3: Plugging the results of Step 1 into Step 2

Differentiating (A3) w.r.t 'x':

$$\frac{dV_c}{dx} = \frac{\left(V_{c2} - V_{c1}\right)}{L} + P\frac{\left(V_{s2} - V_{s1}\right)}{L} - P\frac{dV_s}{dx}$$

and substituting the result in (A5), we can obtain a solution for the charge current as:

$$\begin{aligned} I_C &= -g\left(\frac{(V_{c2}-V_{c1})}{L} + P\frac{(V_{s2}-V_{s1})}{L} - P\frac{dV_s}{dx} + P\frac{dV_s}{dx}\right) \\ &= -g\left(\frac{(V_{c2}-V_{c1})}{L} + P\frac{(V_{s2}-V_{s1})}{L}\right) \\ &= \frac{1}{\rho L}\begin{bmatrix}1 & P\end{bmatrix}\begin{Bmatrix}V_{c1}-V_{c2} \\ V_{s1}-V_{s2}\end{Bmatrix} \end{aligned} \qquad \text{(A6)}$$

Similarly the solution for the spin current is given as

$$\begin{aligned} I_S &= -g\left(P\frac{(V_{c2}-V_{c1})}{L} + P^2\frac{(V_{s2}-V_{s1})}{L} + (1-P^2)\frac{dV_s}{dx}\right) \\ &= \frac{1}{\rho L}\begin{bmatrix}P & P^2\end{bmatrix}\begin{Bmatrix}V_{c1}-V_{c2} \\ V_{s1}-V_{s2}\end{Bmatrix} - \frac{(1-P^2)}{\rho}\frac{dV_s}{dx} \end{aligned}$$

However, (A4) $\Rightarrow$ $\dfrac{dV_s}{dx} = \dfrac{1}{\lambda_{sf}}\dfrac{V_{s2}\cosh\left(\frac{x}{\lambda_{sf}}\right) - V_{s1}\cosh\left(\frac{L-x}{\lambda_{sf}}\right)}{\sinh\left(\frac{L}{\lambda_{sf}}\right)}$

$$\begin{aligned} \therefore I_s\big|_{x=0} = {} & \frac{1}{\rho L}\begin{bmatrix}P & P^2\end{bmatrix}\begin{Bmatrix}V_{c1}-V_{c2} \\ V_{s1}-V_{s2}\end{Bmatrix} + \frac{(1-P^2)}{\rho\lambda_{sf}}\operatorname{cosech}\left(\frac{L}{\lambda_{sf}}\right)\begin{Bmatrix}0 \\ V_{s1}-V_{s2}\end{Bmatrix} \\ & + \frac{(1-P^2)}{\rho\lambda_{sf}}\tanh\left(\frac{L}{2\lambda_{sf}}\right)\begin{Bmatrix}0 \\ V_{s1}\end{Bmatrix} \end{aligned} \qquad \text{(A7)}$$

Grouping (A6) and (A7) we can express the current flowing in the section as

$$\begin{Bmatrix}I_c \\ I_s\end{Bmatrix} = \frac{1}{\rho L}\begin{pmatrix}1 & P \\ P & P^2 + \dfrac{(1-P^2)}{(\lambda_{sf}/L)}\operatorname{cosech}\left(\dfrac{L}{\lambda_{sf}}\right)\end{pmatrix}\begin{Bmatrix}V_{c1}-V_{c2} \\ V_{s1}-V_{s2}\end{Bmatrix} + \begin{pmatrix}0 & 0 \\ 0 & \dfrac{(1-P^2)}{\rho\lambda_{sf}}\tanh\left(\dfrac{L}{2\lambda_{sf}}\right)\end{pmatrix}\begin{Bmatrix}0 \\ V_{s1}\end{Bmatrix}$$

or more concisely in the form of a $\pi$-network of series $(G^{se})$ and shunt $(G^{sh})$ conductance matrices

$$\begin{Bmatrix}I_c \\ I_s\end{Bmatrix}_1 = \left[G^{se}\right]_{2\times 2}\begin{Bmatrix}\Delta V_c \\ \Delta V_s\end{Bmatrix} + \left[G^{sh}\right]_{2\times 2}\begin{Bmatrix}0 \\ V_{s1}\end{Bmatrix}$$

where $G^{se} = \dfrac{1}{\rho L}\begin{pmatrix}1 & p \\ p & p^2 + \alpha\ \text{cosech}\left(\dfrac{L}{\lambda_{sf}}\right)\end{pmatrix}$; $\quad G^{sh} = \dfrac{1}{\rho L}\begin{pmatrix}0 & 0 \\ 0 & \alpha\ \tanh\left(\dfrac{L}{2\lambda_{sf}}\right)\end{pmatrix}$; $\alpha = (1-p^2)\left(\dfrac{L}{\lambda_{sf}}\right)$

### B. Derivation of Non-local resistance of a spin valve (Eq. 2.8)

In this section we will provide a detailed analytical derivation of the expression for the non-local resistance of a spin valve given by Eq. (2.8), i.e.,

$$R_{NL}=\pm 2R_{SN}e^{-L/\lambda_N}\frac{\left(\dfrac{P_{T1}\dfrac{R_{T1}}{R_{SN}}}{1-P_{T1}^2}+\dfrac{P_{f1}\dfrac{R_{F1}}{R_{SN}}}{1-P_{f1}^2}\right)\times\left(\dfrac{P_{T2}\dfrac{R_{T2}}{R_{SN}}}{1-P_{T2}^2}+\dfrac{P_{f2}\dfrac{R_{F2}}{R_{SN}}}{1-P_{f2}^2}\right)}{\left(1+\dfrac{2\dfrac{R_{T1}}{R_{SN}}}{1-P_{T1}^2}+\dfrac{2\dfrac{R_{F1}}{R_{SN}}}{1-P_{f1}^2}\right)\times\left(1+\dfrac{2\dfrac{R_{T2}}{R_{N}}}{1-P_{T2}^2}+\dfrac{2\dfrac{R_{F2}}{R_{SN}}}{1-P_{f2}^2}\right)-e^{-2L/\lambda_N}} \qquad (2.8)$$

The starting point is the conductance matrix $[G]_{ckt}$ for the entire non-local spin valve structure shown in Fig. 8, which can be set up as follows:

$$\begin{array}{c|ccccccc}
 & 1 & 2 & 3 & 4 & 5 & 6 & 7 \\
\hline
1 & G_{0F1}+G_{F1} & -G_{F1} & 0 & 0 & 0 & 0 & 0 \\
2 & -G_{F1} & G_{0F1}+G_{F1}+G_{T1} & 0 & 0 & -G_{T1} & 0 & 0 \\
3 & 0 & 0 & G_{0F2}+G_{F2} & -G_{F2} & 0 & 0 & 0 \\
4 & 0 & 0 & -G_{F2} & G_{0F2}+G_{F2}+G_{T2} & 0 & -G_{T2} & 0 \\
5 & 0 & -G_{T1} & 0 & 0 & G_{N1}+G_{0N1}+G_{N2}+G_{0N2}+G_{T1} & -G_{N2} & 0 \\
6 & 0 & 0 & 0 & -G_{T2} & -G_{N2} & G_{N3}+G_{0N3}+G_{N2}+G_{0N2}+G_{T2} & -G_{N3} \\
7 & 0 & 0 & 0 & 0 & 0 & -G_{N3} & G_{N3}+G_{0N3}
\end{array}$$

The subscripts *F*, *T* and *N* correspond to sections of the ferromagnet, tunnel-barrier and non-magnetic channel respectively, while the $G$ and $G_0$ refer to the 2-component series and shunt conductance matrices describing each of these different sections. The inputs to each of these matrices are the material parameters (resistivity and polarization) and physical dimensions (area, length etc) of these different sections. ***The Matlab code in Appendix C*** computes the various nodal quantities for the entire structure by solving:

$$\begin{Bmatrix}\vec{I}_1\\0\\\vdots\\0\end{Bmatrix}=\left[G_{dev}\right]\begin{Bmatrix}\vec{V}_1\\\vec{V}_2\\\vdots\\\vec{V}_7\end{Bmatrix}$$

The currents corresponding to nodes '2' through '7' are set to zero because of the absence of external current sources. The non-local resistance is then given by $(V_{3c}-V_{7c})/I_{1c}$. One can also compute the non-local resistance analytically from $[G_{ckt}]$ and arrive at Eq. (2.8). In order to simplify such an analysis, it is useful to first make a few simplifications with the 2-component conductance matrices for the different sections that enter $[G_{ckt}]$. For example, a tunnel barrier has a very short length along the transport dimension and consequently the term $L/\lambda_{sf}\to 0$ .Under this condition the tunnel barrier can be represented by

$$\textit{Tunnel Barrier}:\ G_T = g_T\begin{pmatrix} 1 & p_T \\ p_T & 1 \end{pmatrix};\quad G_{0T} = [0];$$

where $g_T$ is the total conductance of the barrier and $p_T$ is its effective spin polarization. Similarly one can assume that the Ferromagnets in the non-local spin valve are characterized by $L/\lambda_{sf} >> 1$ so that the conductance matrices can now be represented by

$$\textit{Ferro-magnet}:\ G_F = g_F\begin{pmatrix} 1 & p_F \\ p_F & p_F^2 \end{pmatrix};\quad G_{0F} = g_F\begin{pmatrix} 0 & 0 \\ 0 & \left(1-p_F^2\right)l_F \end{pmatrix};$$

where $l_F = L/\lambda_{sf}$ and $g_F$ , $p_F$ refer to the conductance and polarization respectively.
In the case of a non-magnetic material, when $L/\lambda_{sf} >> 1$, the above equation becomes

$$\textit{Non-magnetic channel}:\ G_N = g_N\begin{pmatrix} 1 & 0 \\ 0 & 0 \end{pmatrix};\quad G_{0N} = g_N\begin{pmatrix} 0 & 0 \\ 0 & l_N \end{pmatrix};$$

where $g_N$ is the total conductance of the non-magnetic section and $l_N = L/\lambda_{sf}$ .

Keeping these approximations in mind, we can start analyzing $\left[G_{ckt}\right]$ row by row to obtain the voltage at each node in the device.

**<u>Detector side</u>:** The goal is to obtain a relation for the measured voltage i.e. $V_{3c} - V_{7c}$

1) Row 3 of $\left[G_{ckt}\right]$ implies that:

$$\left[G_{0F2} + G_{F2}\right]\begin{Bmatrix} V_{3c} \\ 0 \end{Bmatrix} - \left[G_{F2}\right]\begin{Bmatrix} V_{4c} \\ V_{4s} \end{Bmatrix} = 0$$

$$\Rightarrow\ \underbrace{g_{F2}\begin{pmatrix} 0 & 0 \\ 0 & (1-p_{F2}^2)l_{F2} \end{pmatrix}\begin{Bmatrix} V_{3c} \\ 0 \end{Bmatrix}}_{0} + g_{F2}\begin{pmatrix} 1 & p_{F2} \\ p_{F2} & p_{F2}^2 \end{pmatrix}\begin{Bmatrix} V_{3c} - V_{4c} \\ -V_{4s} \end{Bmatrix} = 0$$

giving the result $\boxed{V_{3c} = V_{4c} + p_{F2}V_{4s}}$ (B1)

2) Row 4 of $\left[G_{ckt}\right]$ implies that:

$$-\left[G_{F2}\right]\begin{Bmatrix} V_{3c} \\ 0 \end{Bmatrix} + \left[G_{0F2} + G_{F2} + G_{T2}\right]\begin{Bmatrix} V_{4c} \\ V_{4s} \end{Bmatrix} - \left[G_{T2}\right]\begin{Bmatrix} V_{6c} \\ V_{6s} \end{Bmatrix} = 0$$

By substituting row 3 into the above equation, we can eliminate $G_{F2}$, which gives

$$\left[G_{0F2}\right]\begin{Bmatrix} V_{4c} \\ V_{4s} \end{Bmatrix} + \left[G_{T2}\right]\begin{Bmatrix} V_{4c} - V_{6c} \\ V_{4s} - V_{6s} \end{Bmatrix} = 0$$

which can then be simplified to give

$$\boxed{V_{3c} = V_{6c} + P_2^{eff}V_{6s} \quad \text{where} \quad P_2^{eff} = \frac{g_{F2}l_{F2}\left(1-p_{F2}^2\right)p_{T2} + g_{T2}\left(1-p_{T2}^2\right)p_{F2}}{g_{F2}l_{F2}\left(1-p_{F2}^2\right) + g_{T2}\left(1-p_{T2}^2\right)}} \quad \text{(B2)}$$

Proof:

$$g_{F2}\begin{pmatrix}0 & 0\\ 0 & (1-p_{F2}^2)l_{F2}\end{pmatrix}\begin{Bmatrix}V_{4c}\\ V_{4s}\end{Bmatrix}+g_{T2}\begin{pmatrix}1 & p_{T2}\\ p_{T2} & 1\end{pmatrix}\begin{Bmatrix}V_{4c}-V_{6c}\\ V_{4s}-V_{6s}\end{Bmatrix}=\begin{Bmatrix}0\\ 0\end{Bmatrix}$$

$$\Rightarrow\left(V_{4c}-V_{6c}\right)+p_{T2}\left(V_{4s}-V_{6s}\right)=0 \qquad \text{(I)}$$

and $g_{F2}(1-p_{F2}^2)l_{F2}V_{4s}+g_{T2}\left\{p_{T2}\left(V_{4c}-V_{6c}\right)+\left(V_{4s}-V_{6s}\right)\right\}=0$ (II)

(I) in (II) $\equiv g_{F2}(1-p_{F2}^2)l_{F2}V_{4s}+g_{T2}\left\{\left(1-p_{T2}^2\right)\left(V_{4s}-V_{6s}\right)\right\}=0$

$$\Rightarrow V_{4s}=\frac{g_{T2}\left(1-p_{T2}^2\right)}{g_{F2}(1-p_{F2}^2)l_{F2}+g_{T2}\left(1-p_{T2}^2\right)}V_{6s} \qquad \text{(III)}$$

(I) $\Rightarrow V_{4c}=V_{6c}-p_{T2}\left(V_{4s}-V_{6s}\right)$

$$=V_{6c}+p_{T2}\frac{g_{F2}(1-p_{F2}^2)l_{F2}}{g_{F2}(1-p_{F2}^2)l_{F2}+g_{T2}\left(1-p_{T2}^2\right)}V_{6s} \qquad \text{(IV)}$$

(III) and (IV) in (A1) leads to (A2)

3) Solving Row 7 gives

$$\left[G_{0N3}\right]\begin{Bmatrix}V_{7c}\\ V_{7s}\end{Bmatrix}+\left[G_{N3}\right]\begin{Bmatrix}V_{7c}-V_{6c}\\ V_{7s}-V_{6s}\end{Bmatrix}=0$$

$$g_{N3}\begin{pmatrix}0 & 0\\ 0 & l_{N3}\end{pmatrix}\begin{Bmatrix}V_{7c}\\ V_{7s}\end{Bmatrix}+g_{N3}\begin{pmatrix}1 & 0\\ 0 & 0\end{pmatrix}\begin{Bmatrix}V_{7c}-V_{6c}\\ V_{7s}-V_{6s}\end{Bmatrix}=0$$

$$\Rightarrow V_{7c}=V_{6c} \qquad \text{(V)}$$

The result (V) can be arrived at by simple inspection of the spin valve. Node 7 physically represents a floating node so there is no charge current flowing in between nodes 6 and 7 This means that $V_{7c}=V_{6c}$ and $V_{7s}=0$.

(V) in (IV) gives the net detected voltage as $\boxed{\left(V_{3c}-V_{7c}\right)=P_2^{eff}V_{6s}}$ (B3)

Let us now see how this detected voltage is related to the injected current.

**Injector Side:**

4) Row 1 of $\left[G_{ckt}\right]$ allows us to relate the charge and spin currents by the relation

$$\left[G_{F1}\right]\begin{Bmatrix}V_{1c}-V_{2c}\\ -V_{2s}\end{Bmatrix}+\left[G_{0F1}\right]\begin{Bmatrix}V_{1c}\\ 0\end{Bmatrix}=\begin{Bmatrix}I_{1c}\\ I_{1s}\end{Bmatrix}$$

$$\Rightarrow \quad g_{F1}\begin{pmatrix} 1 & p_{F1} \\ p_{F1} & p_{F1}^2 \end{pmatrix}\begin{Bmatrix} V_{1c}-V_{2c} \\ -V_{2s} \end{Bmatrix} = \begin{Bmatrix} I_{1c} \\ I_{1s} \end{Bmatrix} \qquad \text{(B4)}$$

$$\Rightarrow I_{1s} = p_{F1} I_{1c} \qquad \text{(B5)}$$

This tells us that the spin current flowing through a long ferromagnet is just the charge current times the polarization of the magnet.

5) Row 2 of $\left[G_{ckt}\right]$ gives the relation that

$$\left[G_{T1}\right]\begin{Bmatrix} V_{2c}-V_{1c} \\ V_{2s} \end{Bmatrix} + \left[G_{0F1}\right]\begin{Bmatrix} V_{2c} \\ V_{2s} \end{Bmatrix} + \left[G_{T1}\right]\begin{Bmatrix} V_{2c}-V_{5c} \\ V_{2s}-V_{5s} \end{Bmatrix} = \begin{Bmatrix} 0 \\ 0 \end{Bmatrix}$$

$$\text{(B4) and (B5)} \Rightarrow \left[G_{T1}\right]\begin{Bmatrix} V_{2c}-V_{5c} \\ V_{2s}-V_{5s} \end{Bmatrix} + \left[G_{0F1}\right]\begin{Bmatrix} V_{2c} \\ V_{2s} \end{Bmatrix} = \begin{Bmatrix} I_{1c} \\ p_{F1} I_{1c} \end{Bmatrix}$$

$$\equiv g_{T1}\begin{pmatrix} 1 & p_{T1} \\ p_{T1} & 1 \end{pmatrix}\begin{Bmatrix} V_{2c}-V_{5c} \\ V_{2s}-V_{5s} \end{Bmatrix} + g_{F1}\begin{pmatrix} 0 & 0 \\ 0 & (1-p_{F1}^2) l_{F1} \end{pmatrix}\begin{Bmatrix} V_{2c} \\ V_{2s} \end{Bmatrix} = \begin{Bmatrix} I_{1c} \\ p_{F1} I_{1c} \end{Bmatrix} \qquad \text{(B6)}$$

which can be solved to obtain the relation

$$V_{2s} = \frac{g'_{T1}}{g'_{T1}+g'_{F1}} V_{5s} + \frac{\left(p_{F1}-p_{T1}\right)}{g'_{T1}+g'_{F1}} I_{1c} \qquad \text{(B7)}$$

where $g'_{T1} = g_{T1}\left(1-p_{T1}^2\right)$ and $g'_{F1} = g_{F1} l_{F1}\left(1-p_{F1}^2\right)$

<u>Proof:</u> We have from (B6):

$$g_{T1}\left(V_{2c}-V_{5c}\right) = I_{1c} - p_{T1}\left(V_{2s}-V_{5s}\right) \qquad \text{(VI)}$$

$$p_{F1} I_{1c} = g_{F1}(1-p_{F1}^2) l_{F1} V_{2s} + p_{T1} g_{T1}\left(V_{2c}-V_{5c}\right) + g_{T1}\left(V_{2s}-V_{5s}\right) \qquad \text{(VII)}$$

$$\text{(VI) in (VII)} \Rightarrow p_{F1} I_{1c} = g'_{F1} V_{2s} + p_{T1} I_{1c} + g'_{T1}\left(V_{2s}-V_{5s}\right) \qquad \text{(same as B7)}$$

6) A similar analysis of Row 5 of $\left[G_{ckt}\right]$ with the (B7) substituted in it gives:

$$I_{1c} P_1^{eff} = \left[g_N l_N \left(1+\tanh\left(l_N/2\right)\right) + g_1^{eff}\right] V_{5s} + \left(V_{5s}-V_{6s}\right) g_N \ \text{cosech}\left(l_N\right) \qquad \text{(B8)}$$

where $P_1^{eff} = \dfrac{g_{F1} l_{F1}\left(1-p_{F1}^2\right) p_{T1} + g_{T1}\left(1-p_{T1}^2\right) p_{F1}}{g_{F1} l_{F1}\left(1-p_{F1}^2\right) + g_{T1}\left(1-p_{T1}^2\right)}$; $\quad g_1^{eff} = \dfrac{g_{F1} l_{F1}\left(1-p_{F1}^2\right) * g_{T1}\left(1-p_{T1}^2\right)}{g_{F1} l_{F1}\left(1-p_{F1}^2\right) + g_{T1}\left(1-p_{T1}^2\right)}$

<u>Proof:</u>

$$\left[G_{N1}+G_{0N1}+G_{N2}+G_{0N2}\right]\begin{Bmatrix} V_{5c} \\ V_{5s} \end{Bmatrix} - \left[G_{N2}\right]\begin{Bmatrix} V_{6c} \\ V_{6s} \end{Bmatrix} = \left[G_{T1}\right]\begin{Bmatrix} V_{2c}-V_{5c} \\ V_{2s}-V_{5s} \end{Bmatrix}$$

Note that: $G_{N1} = g_{N1}\begin{pmatrix}1 & 0\\ 0 & 0\end{pmatrix}$; $\quad G_{0N1} = g_{N1}\begin{pmatrix}0 & 0\\ 0 & l_{N1}\end{pmatrix}$; $\quad \because l_{N1} \gg 1$

But $G_{N2} = g_{N2}\begin{pmatrix}1 & 0\\ 0 & l_{N2}\text{cosech}(l_{N2})\end{pmatrix}$; $\quad G_{0N2} = g_{N2}\begin{pmatrix}0 & 0\\ 0 & l_{N2}\tanh(l_{N2}/2)\end{pmatrix}$; $\quad \because l_{N2} \leq 1$

$$\Rightarrow \left[g_{N1}l_{N1} + g_{N2}l_{N2}\text{cosech}(l_{N2}) + g_{N2}l_{N2}\tanh(l_{N2}/2)\right]V_{5s} -$$
$$g_{N2}l_{N2}\text{cosech}(l_{N2})V_{6s} = \underbrace{p_{F1}I_{1c} - g_{F1}l_{F1}\left(1-p_{F1}^2\right)V_{2s}}_{(B6)}$$

$$g_{N1}l_{N1} = g_{N2}l_{N2} \equiv g_N l_N$$
$$\Rightarrow g_N l_N\left(1+\tanh(l_{N2}/2)\right)V_{5s} + g_N l_N\,\text{cosech}(l_{N2})\left(V_{5s}-V_{6s}\right) = p_{F1}I_{1c} - g'_{F1}V_{2s} \qquad \text{(B9)}$$

(B7) in (B9) leads to (B8).

7) Finally we can couple the quantities from the injector and detector sides by obtaining a relation between $V_{5s}$ and $V_{6s}$ from row 6 of $[G_{ckt}]$ as follows:

$$\left[G_{N2}+G_{0N2}+G_{N3}+G_{0N3}\right]\begin{Bmatrix}V_{6c}\\ V_{6s}\end{Bmatrix} - \left[G_{N2}\right]\begin{Bmatrix}V_{5c}\\ V_{5s}\end{Bmatrix} - \left[G_{N3}\right]\begin{Bmatrix}V_{7c}\\ V_{7s}\end{Bmatrix} = \left[G_{T2}\right]\begin{Bmatrix}V_{4c}-V_{6c}\\ V_{4s}-V_{6s}\end{Bmatrix}$$

Equating the terms corresponding to the spin currents we get

$$g_N l_N\left(1+\tanh(l_{N2}/2)\right)V_{6s} + g_N l_N\,\text{cosech}(l_{N2})\left(V_{6s}-V_{5s}\right) = g_{T2}\left[p_{T2}\left(V_{4c}-V_{6c}\right)+\left(V_{4s}-V_{6s}\right)\right]$$
$$= g_{F2}l_{F2}\left(1-p_{F2}^2\right)V_{4s} \text{ from Row 4}$$
$$= g_2^{eff}V_{6s} \text{ from (III)} \qquad \text{(B10)}$$

where $g_2^{eff} = \dfrac{g_{F2}l_{F2}\left(1-p_{F2}^2\right) * g_{T2}\left(1-p_{T2}^2\right)}{g_{F2}l_{F2}\left(1-p_{F2}^2\right) + g_{T2}\left(1-p_{T2}^2\right)}$. Defining $y_2 = g_2^{eff}/\left(g_N l_N\right)$, (B10) can be rearranged as:

$$V_{6s} = V_{5s}\frac{1}{s+c+y_2 s}; \quad s \equiv \sinh(l_N);\ c \equiv \cosh(l_N) \qquad \text{(B11)}.$$

(B11) in (B8) gives

$$I_{1c}P_1^{eff} = V_{5s}\left(g_1^{eff} + g_N l_N\frac{c+s-1}{s} + g_N l_N\frac{c+s+y_2 s-1}{s\left(c+s+y_2 s\right)}\right)$$
$$= V_{6s}\left(c+s+y_2 s\right)\left(g_1^{eff} + g_N l_N\frac{c+s-1}{s} + g_N l_N\frac{c+s+y_2 s-1}{s\left(c+s+y_2 s\right)}\right) \qquad \text{(B12)}$$

**Non-local Resistance:** We can now obtain the non-local resistance by combining the expressions obtained for the injector (B3) and detector side(B12):

$$\frac{(V_{3c}-V_{7c})}{I_{1c}} = \frac{1}{g_N l_N}\frac{P_2^{eff}}{c+s+y_2 s}\frac{P_1^{eff}}{\left(g_1^{eff}/g_N l_N + \frac{c+s-1}{s} + \frac{c+s+y_2 s-1}{s(c+s+y_2 s)}\right)}$$
$$= R_{SN}\frac{P_1^{eff}P_2^{eff}s}{(c+s+y_1 s)(c+s+y_2 s)-1} \tag{B13}$$

where $R_{SN}=1/g_N l_N$ is the resistance of the channel material over one spin diffusion length; and $y_1 \equiv g_1^{eff}/(g_N l_N)$ .

Simplifying the above expression gives

$$\frac{(V_{3c}-V_{7c})}{I_{1c}} = R_{SN}\frac{P_1^{eff}P_2^{eff}s}{e^{2l_{N2}}+(y_1+y_2)se^{l_{N2}}+y_1y_2s^2-1}$$
$$= R_{SN}\frac{P_1^{eff}P_2^{eff}}{(e^{2l_{N2}}-1)/s+(y_1+y_2)e^{l_{N2}}+y_1y_2s}$$
$$= R_{SN}\frac{P_1^{eff}P_2^{eff}}{e^{l_{N2}}\left(y_1+y_2+{y_1y_2}/{2}+2\right)-e^{-l_{N2}}\,{y_1y_2}/{2}}$$
$$\boxed{\therefore \frac{(V_{3c}-V_{7c})}{I_{1c}} = R_{SN}\frac{2P_1^{eff}P_2^{eff}e^{-l_{N2}}}{(y_1+2)(y_2+2)-y_1y_2e^{-2l_{N2}}}}$$

The above expression for the magnetoresistance is a concise representation of Eq. (2.8). Expanding the various terms in the equation we get:

$$\frac{(V_{3c}-V_{7c})}{I_{1c}} = R_{SN}\frac{2e^{-l_{N2}}P_1^{eff}P_2^{eff}/y_1y_2}{(1+2/y_1)(1+2/y_2)-e^{-2l_{N2}}}$$

$$= R_{SN}\frac{2e^{-l_{N2}}\dfrac{\left[\dfrac{g_{F1}l_{F1}(1-p_{F1}^2)p_{T1}+g_{T1}(1-p_{T1}^2)p_{F1}}{g_{F1}l_{F1}(1-p_{F1}^2)+g_{T1}(1-p_{T1}^2)}\right](g_Nl_N)}{\left[\dfrac{g_{F1}l_{F1}(1-p_{F1}^2)*g_{T1}(1-p_{T1}^2)}{g_{F1}l_{F1}(1-p_{F1}^2)+g_{T1}(1-p_{T1}^2)}\right]}\dfrac{\left[\dfrac{g_{F2}l_{F2}(1-p_{F2}^2)p_{T2}+g_{T2}(1-p_{T2}^2)p_{F2}}{g_{F2}l_{F2}(1-p_{F2}^2)+g_{T2}(1-p_{T2}^2)}\right](g_Nl_N)}{\left[\dfrac{g_{F2}l_{F2}(1-p_{F2}^2)*g_{T2}(1-p_{T2}^2)}{g_{F2}l_{F2}(1-p_{F2}^2)+g_{T2}(1-p_{T2}^2)}\right]}}{\left(1+\dfrac{2(g_Nl_N)}{\left[\dfrac{g_{F1}l_{F1}(1-p_{F1}^2)*g_{T1}(1-p_{T1}^2)}{g_{F1}l_{F1}(1-p_{F1}^2)+g_{T1}(1-p_{T1}^2)}\right]}\right)\left(1+\dfrac{2(g_Nl_N)}{\left[\dfrac{g_{F2}l_{F2}(1-p_{F2}^2)*g_{T2}(1-p_{T2}^2)}{g_{F2}l_{F2}(1-p_{F2}^2)+g_{T2}(1-p_{T2}^2)}\right]}\right)-e^{-2l_{N2}}}$$

A straightforward simplification of this expression leads to the Non-local resistance given by

$$\frac{\left(V_{3c}-V_{7c}\right)}{I_{1c}} = R_{SN}\frac{2e^{-l_{N2}}\left(\frac{p_{T1}}{g_{T1}\left(1-p_{T1}^2\right)R_{SN}}+\frac{p_{F1}}{g_{F1}l_{F1}\left(1-p_{F1}^2\right)R_{SN}}\right)\left(\frac{p_{T2}}{g_{T2}\left(1-p_{T2}^2\right)R_{SN}}+\frac{p_{F2}}{g_{F2}l_{F2}\left(1-p_{F2}^2\right)R_{SN}}\right)}{\left(1+\frac{2}{g_{T1}\left(1-p_{T1}^2\right)R_{SN}}+\frac{2}{g_{F1}l_{F1}\left(1-p_{F1}^2\right)R_{SN}}\right)\left(1+\frac{2}{g_{T2}\left(1-p_{T2}^2\right)R_{SN}}+\frac{2}{g_{F2}l_{F2}\left(1-p_{F2}^2\right)R_{SN}}\right)-e^{-2l_{N2}}}$$

i.e

$$R_{NL}=\frac{\left(V_{3c}-V_{7c}\right)}{I_{1c}} = \pm 2R_{SN}e^{-L/\lambda_N}\frac{\left(\frac{P_{T1}\frac{R_{T1}}{R_{SN}}}{1-P_{T1}^2}+\frac{P_{f1}\frac{R_{F1}}{R_{SN}}}{1-P_{f1}^2}\right)\times\left(\frac{P_{T2}\frac{R_{T2}}{R_{SN}}}{1-P_{T2}^2}+\frac{P_{f2}\frac{R_{F2}}{R_{SN}}}{1-P_{f2}^2}\right)}{\left(1+\frac{2\frac{R_{T1}}{R_{SN}}}{1-P_{T1}^2}+\frac{2\frac{R_{F1}}{R_{SN}}}{1-P_{f1}^2}\right)\times\left(1+\frac{2\frac{R_{T2}}{R_N}}{1-P_{T2}^2}+\frac{2\frac{R_{F2}}{R_{SN}}}{1-P_{f2}^2}\right)-e^{-2L/\lambda_N}}$$

.

**C. Derivation of 4-Component Interface Conductance Matrix (Eq. 2.11)**

The conductance matrices in Eq. (2.11) describing the interface of the 'z'-directed ferromagnet (FM) and the non-magnetic channel (NM) were shown to be

$$G_{Int}^{se} = \frac{q^2}{h} M \begin{pmatrix} 1 & P & 0 & 0 \\ P & 1 & 0 & 0 \\ 0 & 0 & 0 & 0 \\ 0 & 0 & 0 & 0 \end{pmatrix} \begin{vmatrix} c \\ z \\ x \\ y \end{vmatrix}; \quad G_{Int}^{sh} = \frac{q^2}{h} M \begin{pmatrix} 0 & 0 & 0 & 0 \\ 0 & 0 & 0 & 0 \\ 0 & 0 & a & b \\ 0 & 0 & -b & a \end{pmatrix} \begin{vmatrix} c \\ z \\ x \\ y \end{vmatrix} \qquad (2.11)$$

where:

$$\frac{q^2}{h} Ma \equiv \sum_j^M \left[ \left(1 - r_{\downarrow} r_{\uparrow}^{*}\right) + \left(1 - r_{\uparrow} r_{\downarrow}^{*}\right) \right] \Big/ 2 \quad \text{and} \quad \frac{q^2}{h} Mb \equiv \sum_j^M i \left[ \left(1 - r_{\downarrow} r_{\uparrow}^{*}\right) - \left(1 - r_{\uparrow} r_{\downarrow}^{*}\right) \right] \Big/ 2$$

These matrices can be derived from scattering theory and the details of this derivation are listed below. The principle here is that one can calculate the transmission and reflection probabilities for an electron wavefunction incident from a plane inside the NM to a plane inside the FM as shown in the Fig. C.1.

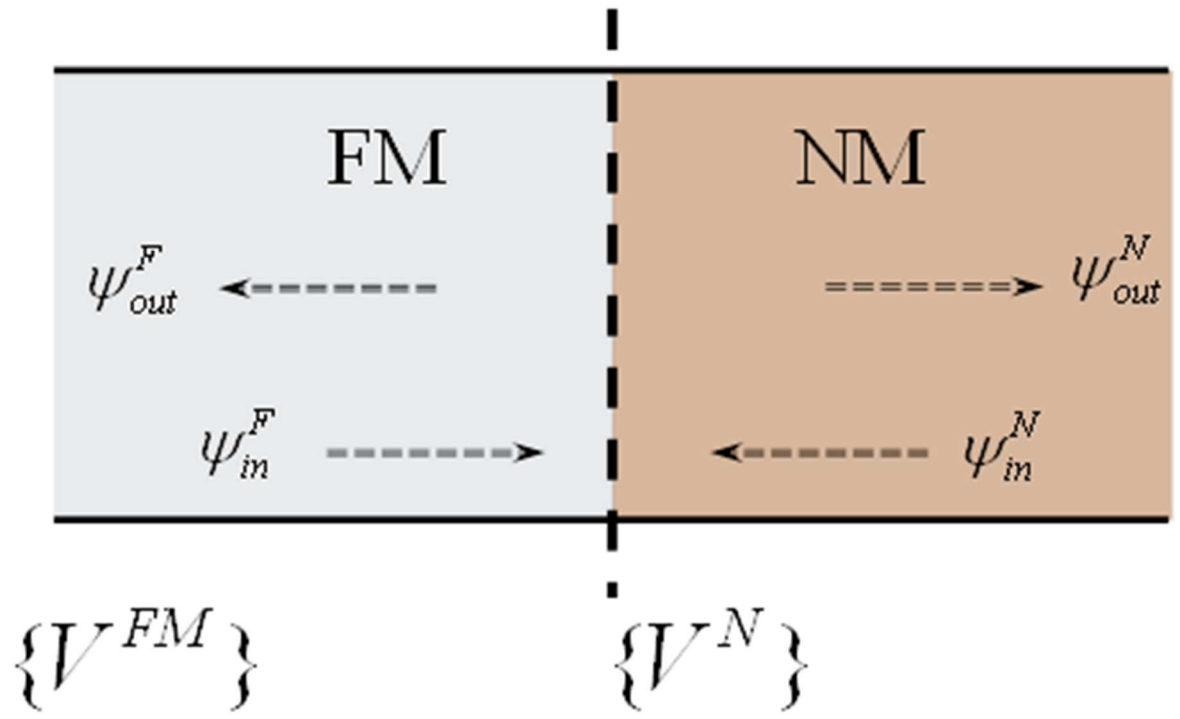

Fig. C.1. Scattering theory description for transport across a FM-NM interface

The relationship between the incoming and outgoing wave functions can be written in terms of transmission (τ) and reflection (ρ) coefficients as:

$$\begin{Bmatrix} \psi_{out}^{N} \\ \psi_{out}^{F} \end{Bmatrix} = \begin{bmatrix} \rho & \tau \\ \tau' & \rho' \end{bmatrix} \begin{Bmatrix} \psi_{in}^{N} \\ \psi_{in}^{F} \end{Bmatrix}$$

The electron current flowing back into the NM region can be thought of as

$$i_{out}^{N} \approx \frac{q}{h} (\psi_{out}^{N} \; \psi_{out}^{N\dagger}) = \frac{q}{h} \left( \rho \psi_{in}^{N} \; \psi_{in}^{N\dagger} \rho^{\dagger} + \tau \psi_{in}^{F} \; \psi_{in}^{F\dagger} \tau^{\dagger} \right)$$

$$= \frac{q}{h} \left( \rho f^{N} \rho^{\dagger} + \tau f^{F} \tau^{\dagger} \right)$$

The net current flowing across the interface from the NM to FM region is then given by

$$\boxed{\begin{aligned} i &= i^N_{out} - i^N_{in} \\ &= \frac{q}{h}\Big[ - \underbrace{\left(f^N - \rho f^N \rho^\dagger\right)}_{\text{I}} + \underbrace{\tau f^F \tau^\dagger}_{\text{II}} \Big] \end{aligned}} \qquad \text{(C1)}$$

In order to obtain the matrices in Eq. (2.11), we will now show how Eq. (C1) can be represented in the form

$$\boxed{\{\vec{i}\} = G^{se}_{int}\left(\{V^N\} - \{V^F\}\right) + G^{sh}_{int}\{V^N\}}$$

The '$f$' in Eq.(C1) is equivalent to the quasi Fermi level for the electrons in the FM and NM regions. In general, the spin current in the NM regions can be in any direction (non-collinear) resulting in a "vector" form of $f$ which can be then represented by 2x2 matrix in density matrix representation:

$$f^N = \begin{pmatrix} f^N_c + f^N_{sz} & f^N_{sx} - if^N_{sy} \\ f^N_{sx} + if^N_{sy} & f^N_c - f^N_{sz} \end{pmatrix} = \left(f^N_c I + \vec{\sigma}\bullet\vec{f}^N_s\right)$$

For the FM region, we assume that the FM is in the "z" direction and hence:

$$f^F = \left(f^F_c I + \sigma_z f^F_s\right)$$

The τ and ρ are also 2x2 matrices since they also include the spin nature of the electron. For FM pointing in the 'z' direction, these coefficients can be given by

$$\tau = \begin{pmatrix} t_\uparrow & 0 \\ 0 & t_\downarrow \end{pmatrix} \equiv \tau_c I + \sigma_z \tau_s \ ; \qquad \rho = \begin{pmatrix} r_\uparrow & 0 \\ 0 & r_\downarrow \end{pmatrix} \equiv \rho_c I + \sigma_z \rho_s \qquad \text{(C2)}$$

where $r_{\uparrow,\downarrow}$ and $t_{\uparrow,\downarrow}$ are the reflection and transmission coefficients respectively. In Eq. (C2) the charge component of ρ is given by $\rho_c = \left(r_\uparrow + r_\downarrow\right)/2$ while the spin component is given by $\rho_s = \left(r_\uparrow - r_\downarrow\right)/2$. Using these transformations, one can then calculate the different components of the net current flowing across the interface from Eq. C1 as

$$\begin{aligned} \text{I} &\equiv \left(f^N - \rho f^N \rho^\dagger\right) \\ &= \left(f^N_c I + \vec{\sigma}\bullet\vec{f}^N_s\right) - \left(\rho_c I + \sigma_z \rho_s\right)\left(f^N_c I + \vec{\sigma}\bullet\vec{f}^N_s\right)\left(\rho^*_c I + \sigma_z \rho^*_s\right) \end{aligned}$$

and

$$\text{II} \equiv \tau f^F \tau^\dagger = \left(\tau_c I + \sigma_z \tau_s\right)\left(f^F_c I + \sigma_z f^F_s\right)\left(\tau^*_c I + \sigma_z \tau^*_s\right) \qquad \text{(C3)}$$

The rest of this derivation simply relates to the simplification of Eq. (C3) to obtain the conductance in terms of the reflection and transmission coefficients. To do so:

1) We first define some convenient parameters *R, R', T, P* and *Q* such that

$$R \equiv (r_\uparrow r_\uparrow^* + r_\downarrow r_\downarrow^*)/2; \qquad R' \equiv (r_\downarrow r_\uparrow^* + r_\uparrow r_\downarrow^*)/2$$

so that $$\rho_c\rho_c^* = \frac{\left(r_\uparrow r_\uparrow^* + r_\downarrow r_\downarrow^*\right) + \left(r_\downarrow r_\uparrow^* + r_\uparrow r_\downarrow^*\right)}{4} \equiv \frac{R+R'}{2} \quad \text{and} \quad \vec{\rho}_s\vec{\rho}_s^* \equiv \frac{R-R'}{2} \qquad \text{(C4)}$$

this leads to the relation

$$\rho_c\rho_c^* + \rho_s\rho_s^* = R \;; \quad \text{We can similarly define}$$

$$\tau_c\tau_c^* + \tau_s\tau_s^* = T \quad \ni \quad R + T = 1$$

Define $$P \equiv \left(r_\uparrow r_\uparrow^* - r_\downarrow r_\downarrow^*\right)/2; \qquad iQ \equiv \left(r_\downarrow r_\uparrow^* - r_\uparrow r_\downarrow^*\right)/2$$

$$\Rightarrow \rho_c\rho_s^* = \hat{z}\frac{\left(r_\uparrow r_\uparrow^* - r_\downarrow r_\downarrow^*\right) + \left(r_\downarrow r_\uparrow^* - r_\uparrow r_\downarrow^*\right)}{4} = \hat{z}\frac{(P+iQ)}{2} \qquad \text{(C5)}$$

this leads to the relation $\rho_c\rho_s^* + \rho_c^*\rho_s = P \;; \quad \rho_c\rho_s^* - \rho_c^*\rho_s = iQ$

2) Let us look at the first term 'I' in Eq. (C3)

$$f^N \equiv \left(f_c^N I + \vec{\sigma}\bullet\vec{f}_s^N\right) \qquad \text{(C6)}$$

And

$$\rho f^N \rho^\dagger = \left(\rho_c I + \sigma_z\rho_s\right)\left(f_c^N I + \vec{\sigma}\bullet\vec{f}_s^N\right)\left(\rho_c^* I + \sigma_z\rho_s^*\right)$$

$$= (\rho_c a + \rho_s b) + \sigma_z\left(a\rho_s + b\rho_c\right) + (\sigma_x f_{sx}^N + \sigma_y f_{sy}^N)(\rho_c\rho_c^* - \rho_s\rho_s^*) + i(\rho_c\rho_s^* - \rho_s\rho_c^*)(\sigma_x f_{sy}^N - \sigma_y f_{sx}^N)$$

Where we define:

$$a = \rho_c^* f_c^N + \rho_s^* f_{sz}^N; \qquad b = \rho_c^* f_{sz}^N + \rho_s^* f_c^N$$

Then it can be further simplified by noting: $f_{sx}^N\hat{x} + f_{sy}^N\hat{y} = \left(\hat{z}\times\vec{f}_s^N\times\hat{z}\right)$

$$\rho f^N\rho^\dagger = (\rho_c a + \rho_s b) + \sigma_z\left(a\rho_s + b\rho_c\right) - \vec{\sigma}\cdot(\hat{z}\times\hat{z}\times\vec{f}_s^N)R' + Q\vec{\sigma}\cdot(\hat{z}\times\vec{f}_s^N) \qquad \text{(C7)}$$

Grouping (C6) and (C7):

$$f^N - \rho f^N\rho^\dagger = \left(f_c^N - \rho_c a - \rho_s b\right)I + \vec{\sigma}\bullet\left(\vec{f}_s^N - a\rho_s\hat{z} - b\rho_c\hat{z} + (\hat{z}\times\hat{z}\times\vec{f}_s^N)R' - Q(\hat{z}\times\vec{f}_s^N)\right) \qquad \text{(C8)}$$

Similarly, the contribution from the term 'II' in Eq. (C3) can also be extracted as

$$\tau f^F\tau^\dagger = \left(f_c^F T - Pf_{sz}^F\right)I + \sigma_z\left(f_s^F T - Pf_c^F\right)$$

(C9)

The current operator, (C1), is then:

$$I_{op} = \frac{q}{h}\left[\left(f_c^F T - Pf_{sz}^F\right)I - \left(f_c^N - \rho_c a - \rho_s b\right)I\right]$$
$$+ \frac{q}{h}\vec{\sigma}\bullet\left[(f_s^F T - Pf_c^F)\hat{z} - \left(\vec{f}_s^N - a\rho_s\hat{z} - b\rho_c\hat{z} + (\hat{z}\times\hat{z}\times\vec{f}_s^N)R' - Q(\hat{z}\times\vec{f}_s^N)\right)\right] \qquad \text{(C10)}$$

3) <u>*Charge Currents:*</u>

The net charge current can be obtained by:

$$I_c = \text{Trace}(I_{op}) = 2\frac{q}{h}\left\{T\left(f_c^F - f_c^N\right) - \left(f_{sz}^F - f_{sz}^N\right)P\right\} \qquad \text{(C11)}$$

Here we notice that: $f_c^N - \rho_c a - \rho_s \bullet b = (f_c^N T - f_{sz}^N P)I$

The factor of 2 enters Eq. (C11) due to taking trace over identity matrix.

4) _Spin currents:_

The spin current can be obtained by:

$$\vec{I}_S = \text{Trace}(I_{op}\vec{\sigma})$$
$$= \frac{q}{h}\left[-2P\left(f_c^F - f_c^N\right)\hat{z} + 2T\left(f_{sz}^F - f_{sz}^N\right) + 2Q\left(\hat{z}\times\vec{f}_s^N\right) + 2\left(1-R'\right)\left(\hat{z}\times\hat{z}\times\vec{f}_s^N\right)\right] \quad \text{(C12)}$$

Here we use:

$$\left(\vec{f}_s^N - a\rho_s\hat{z} - b\rho_c\hat{z} + (\hat{z}\times\hat{z}\times\vec{f}_s^N)R' - Q(\hat{z}\times\vec{f}_s^N)\right)$$
$$= f_{sz}^N\left(1-R\right)\hat{z} - f_c^N P\hat{z} - Q\left(\hat{z}\times\vec{f}_s^N\right) - \left(1-R'\right)\left(\hat{z}\times\hat{z}\times\vec{f}_s^N\right)$$

5) The overall currents can be grouped together from equations C(11-12) as:

$$I_C = \frac{q}{h}\left(2T\left(f_c^F - f_c^N\right) - 2P\left(f_{sz}^F - f_{sz}^N\right)\right)$$
$$\vec{I}_S = \frac{q}{h}\left[-2P\left(f_c^F - f_c^N\right)\hat{z} + 2T\left(f_{sz}^F - f_{sz}^N\right) + 2Q\left(\hat{z}\times\vec{f}_s^N\right) + 2\left(1-R'\right)\left(\hat{z}\times\hat{z}\times\vec{f}_s^N\right)\right]$$

We would like to rewrite currents in terms of voltages. The above expressions are for single energy level. We have to take into account all the energies involved in the transport by integrating over the energy:

$$I_C = \frac{q}{h}\int\left(2T\left(f_c^F - f_c^N\right) - 2P\left(\vec{f}_s^F - \vec{f}_s^N\right)\bullet\hat{z}\right)dE$$
$$= \frac{q}{h}\int\left[2T\left(-\frac{\partial f}{\partial E}(\mu_c^F - \mu_c^N)\right) - 2P\left(-\frac{\partial f}{\partial E}(\vec{\mu}_s^F - \vec{\mu}_s^N)\right)\bullet\hat{z}\right]dE$$

Using $-qV = \mu$, and at low temperature, the charge current:

$$I_C = \frac{q^2}{h}\left(2T\left(V_c^N - V_c^F\right) - 2P\left(\vec{V}_s^N - \vec{V}_s^F\right)\bullet\hat{z}\right) \quad \text{(C13)}$$

Similarly, noting that at equilibrium, the occupation factor $\vec{f}_s^0 = 0$

$$\vec{f}_s^N = \vec{f}_s^N - \vec{f}_s^0 = -\frac{\partial f}{\partial E}\vec{\mu}_s^N$$

Then the spin current can be also written in form of voltage as:

$$\vec{I}_S = \frac{q^2}{h}\left[-2P\left(V_c^N - V_c^F\right)\hat{z} + 2T\left(V_{sz}^N - V_{sz}^F\right) - 2Q\left(\hat{z}\times\vec{V}_s^N\right) - 2\left(1-R'\right)\left(\hat{z}\times\hat{z}\times\vec{V}_s^N\right)\right] \quad \text{(C14)}$$

This is the current per conducting mode. To compute the total current, the above equation should be summed over all conducting modes. Here we also assume that all the modes are decoupled and it leads to the following simplifications:

i. The quantity $T = \left(t_\uparrow t_\uparrow^* + t_\downarrow t_\downarrow^*\right)/2$, which is the average of the transmission probabilities of the up and down spins into the FM region for a single mode. When $T$ is integrated over all the conducting modes it results in the net interface conductance due to the up and down spins, i.e., $\frac{q^2}{h}\bar{T} = (g_\uparrow + g_\downarrow)/2$. Therefore, we can write down $2\frac{q^2}{h}\bar{T} \equiv g$ where $g$ is the interface conductance.

ii. Similarly, after summing over modes $-P = -(r_\uparrow r_\uparrow^* - r_\downarrow r_\downarrow^*)/2 \rightarrow -\bar{P} = (g_\uparrow - g_\downarrow)/2$. This can be expressed in terms of the interface conductance '$g$' by noting that $(g_\uparrow - g_\downarrow) = \frac{(g_\uparrow - g_\downarrow)}{(g_\uparrow + g_\downarrow)}(g_\uparrow + g_\downarrow) \equiv pg$ where '$p$' is the interface polarization, giving us the result: $-2\frac{q^2}{h}\bar{P} \equiv pg$.

Using the simplifications (i.) and (ii.), the equations for current (C13) and (C14) (after summing over modes) can now be concisely expressed in the matrix form as

$$\begin{Bmatrix} I_c \\ I_z \\ I_x \\ I_y \end{Bmatrix} = \begin{pmatrix} g & pg & 0 & 0 \\ pg & g & 0 & 0 \\ 0 & 0 & \sum_j^M 2(1-R_j') & \sum_j^M 2Q_j \\ 0 & 0 & -\sum_j^M 2Q_j & \sum_j^M 2(1-R_j') \end{pmatrix} \begin{Bmatrix} \Delta V_c \\ \Delta V_z \\ V_x^N \\ V_y^N \end{Bmatrix}$$

The above matrix form is equivalent to Eq. (102) and (103) of [18] by noting:

$$\vec{m} = \hat{z};$$

$$G^\uparrow + G^\downarrow = g; G^\uparrow - G^\downarrow = pg$$

$$\sum_j^M (1-R_j') = \mathrm{Re}\, G^{\uparrow\downarrow}; \ \sum_j^M Q_j = \mathrm{Im}\, G^{\uparrow\downarrow}$$

For a ballistic interface, we can assume that the interface conductance is $g \approx \frac{q^2}{h}M$ where M is the total number of conducting modes at the interface. The matrix can now be split into series and shunt sections as in:

$$\boxed{\{\vec{i}\} = G_{int}^{se}\left(\{V^N\} - \{V^F\}\right) + G_{int}^{sh}\{V^N\}}$$

Here

$$\boxed{G_{Int}^{se} = \frac{q^2}{h}M \begin{pmatrix} 1 & P & 0 & 0 \\ P & 1 & 0 & 0 \\ 0 & 0 & 0 & 0 \\ 0 & 0 & 0 & 0 \end{pmatrix}; \quad G_{Int}^{sh} = \frac{q^2}{h}M \begin{pmatrix} 0 & 0 & 0 & 0 \\ 0 & 0 & 0 & 0 \\ 0 & 0 & a & b \\ 0 & 0 & -b & a \end{pmatrix}}$$

where $\boxed{\frac{q^2}{h}Ma \equiv \sum_j^M (1-R_j') \text{ and } \frac{q^2}{h}Mb \equiv \sum_j^M Q_j}$

## D. Matlab scripts

```
% Example 1: Spin circuit description of a non-local spin valve
% Srikant Srinivasan, Supriyo Datta, Purdue University
clear all

%Constants (all MKS, except energy which is in eV)
q=1.6e-19;Z=zeros(2,2);

% Parameters
% note: R => rho*lambda_sf/A; L => L/lambda_sf
ii=0;
for X=-5:0.1:5
    ii=ii+1; RT(ii)=X; RT1=10^X; RT2=10^X;% Tunnel Resistance
    PT1=0.2; PT2=0.2;% Polarization of Tunnel Contacts
    RF=1e-2; LF1=100; LF2=100; PF1=0.05; PF2=0.05;% Ferromagnetic contacts
    RN=1; LN1=100; LN2=1e-3; LN3=100;% Nonmagnetic Channel

    % Ferromagnetic contacts
    GF1 = ((1/RF/LF1)*[1 PF1;PF1 PF1*PF1])+(((1-PF1*PF1)/RF)*[0 0; 0 ...
                csch(LF1)]);
    GF2 = ((1/RF/LF2)*[1 PF2;PF2 PF2*PF2])+(((1-PF2*PF2)/RF)*[0 0; 0 ...
                csch(LF2)]);
    G0F1 = ((1-PF1*PF1)/RF)*[0 0;0 coth(LF1)-csch(LF1)];
    G0F2 = ((1-PF2*PF2)/RF)*[0 0;0 coth(LF2)-csch(LF2)];

    % Normal channel
    GN1 = (1/RN/LN1)*[1 0;0 LN1*csch(LN1)];
    GN2 = (1/RN/LN2)*[1 0;0 LN2*csch(LN2)];
    GN3 = (1/RN/LN3)*[1 0;0 LN3*csch(LN3)];
    G0N1 = (1/RN)*[0 0;0 coth(LN1)-csch(LN1)];
    G0N2 = (1/RN)*[0 0;0 coth(LN2)-csch(LN2)];
    G0N3 = (1/RN)*[0 0;0 coth(LN3)-csch(LN3)];

    % Tunnel resistances
    GT1=(1/RT1)*[1 PT1;PT1 1];
    GT2=(1/RT2)*[1 PT2;PT2 1];

    % Conductance matrix from KCL
    G = [G0F1+GF1 -GF1 Z Z Z Z Z;
        -GF1 G0F1+GF1+GT1 Z Z -GT1 Z Z;
        Z Z G0F2+GF2 -GF2 Z Z Z;
        Z Z -GF2 G0F2+GF2+GT2 Z -GT2 Z;
        Z -GT1 Z Z GN1+G0N1+G0N2+GN2+GT1 -GN2 Z;
        Z Z Z -GT2 -GN2 GN3+G0N2+G0N3+GN2+GT2 -GN3;
        Z Z Z Z Z -GN3 GN3+G0N3];

    C = [1; PF1; zeros(12,1)];%  Terminal currents
```

```
    V=G\C; V=reshape(V,2,7);% Terminal voltages
    Vout(ii)=V(1,3)-V(1,7);% Output voltage

    %%% Takahashi and Maekawa formula (PRB. 67, 052409)
    RF1=RF; RF2=RF;
    Numer = 2*RN*exp(-LN2)*(PT1*RT1/RN/(1-PT1^2) + PF1*RF1/RN/(1-PF1^2))...
        *(PT2*RT2/RN/(1-PT2^2) + PF2*RF2/RN/(1-PF2^2));
    denom = (1+ 2*RT1/RN/(1-PT1^2) + 2*RF1/RN/(1-PF1^2))...
        *(1+ 2*RT2/RN/(1-PT2^2) + 2*RF2/RN/(1-PF2^2)) - exp(-LN2);
    Rnon_local(ii)=Numer/denom;
end

hold on
plot(RT,Vout/C(1),'r*');
plot(RT,Rnon_local,'bo');
set(gca,'linewidth',[3.0]);
set(gca,'Fontsize',[24]);
xlabel(' log10(Tunnel resistance) -->')
ylabel(' Output Voltage -->')
grid on
```

```
%%% Example 2: Simple LLG solver to reproduce Fig. 16
%%%  Behtash Behin-Aein, Angik Sarkar, Srikant Srinivasan, Vinh Diep,
%%%  Supriyo Datta Research group, Purdue University (2010)

clear all; clc
global hext hd alpha Is_conv

%%% LLG parameters
%%%%%%%%%%%%%%%%%%%%%%%%%%%%%%

%%% Constants
%%%------------
q=1.6e-19; % Coulombs
hbar=6.626e-34/2/pi; % Reduced Planck's constant (J-s)
mub=9.274e-21; % Bohr Magneton
alpha = 0.007; % Gilbert damping parameter
g = 1.76e7; % Gyromagnetic ratio [(rad)/(Oe.s)]

%%% Magnet Parameters (taken from experiment)
%%%-------------------------------------------
Ms = 780; % Saturation Magnetization [emu/cm^3]
Ku2 = 3.14e4; % Uni. anisotropy constant [erg/cm^3]
V = (170*80*2)*1e-21; % Volume [cm^3] ()
Hk = 2*Ku2/Ms ; % Switching field [Oe]
Hd = 4*pi*Ms; % Demagnetizing field [Oe]
Ns=Ms*V/mub % Number of spins in the magnet

%%% Converting magnet parameters into dimensionless quantities. Note that
%%% in this code we transform the LLG equation into a dimensionless
%%% equation by normalizing it to the time constant 1/(g*Hk).
hk = 1; % dimensionless uniaxial field
hd = Hd/Hk; % dimensionless demag field
hext=0; % Assume no external applied fields
tau_c = (1+alpha^2)/(g*Hk); % LLG time constant

% Conversion factor for Ampere spin current into dimensionless input in
% LLG. The factor below is for the term Is/(q*Ns*g*Hk), noting that
% g=2muB/hbar.
I_H_conv = hbar/2/q/(Ms*V*Hk*1e-7);
Isc = alpha*(1 + hd/2) * (Hk*Ms*V) * 1e-7 * 2*q/hbar;
% Isc = Estimated ampere spin current required for easy axis switching


Is=-1.3*Isc; % Spin current (Amps) incident on magnet
% Is=-3*Isc; %Is=-2*Isc;
Is_conv=Is*I_H_conv;
% switching_time=2*q*Ns/Is; %% Estimated switching time.

%%% Initial conditions of the simulation
mz=0.999; % Magnet slightly off easy axis due to, say, thermal noise
m=[sqrt(1-mz^2) 0 mz]; %Magnet in the x-z plane
```

```
%%%%%%%%%%% Solving the LLG equation
options = odeset('RelTol',1e-8,'AbsTol',1e-9);
NanoS = 15; %% Duration in units of nano-seconds
t_span= [0 NanoS*1e-9]/tau_c; %% Dimensionless time span
[t,x]= ode113('LLGsolver_example2', t_span, m, options);

%%%%%%%%%% Plotting
figure(1)
hold on
%plot(t*tau_c/1e-9,x(:,1),'k-'); % m_x
%plot(t*tau_c/1e-9,x(:,2),'r-'); % m_y
h=plot(t*tau_c/1e-9,x(:,3),'r-'); % m_z
axis([0 15 -1 1])
set(h,'linewidth',3.0)
set(gca,'Fontsize',30)
xlabel(' Time (ns) ')
ylabel(' m_z')

%%%%%%%%%%%%%%%%%%%%%%%%%%%%%%%%%%%%%%%%%%%%%



%%%%%%%%%%%%%%%%%%%%%%%%%%%%%%%%%%%%%%%%%%%%%
function dmdt = LLGsolver_example2(t,m)
% Vinh Diep, Srikant Srinivasan, Deepanjan Datta, Supriyo Datta Research group, Purdue University
(2010)
global hd alpha Is_conv

H=[0*m(1) -hd*m(2) m(3)];   % Internal fields i.e. uniaxial (along z) and demag(along x)

Is1=Is_conv*[0 0 1];

%%% Differential Equation for magnetization Dynamics
dmdt0=(-cross(m,H)-alpha*cross(m,cross(m,H))...
+cross(m,cross(Is1,m))+alpha*cross(m,Is1)) ;

dmdt=dmdt0';
end
%%%%%%%%%%%%%%%%%%%%%%%%%%%%%%%%%%%%%%%%%%%%%
```

```
% Example 3: Wrapper Code for reproducing the X and the Y axis of Fig. 18
% Assume that the injector magnet is a fixed layer along a reference
% direction 'z', which also corresponds to the transport direction. The
% detector magnet is a free layer and is initially slightly away from 'z'
% by a few degrees.
% Angik Sarkar, Behtash Behin-Aein, Srikant Srinivasan,
% Supriyo Datta Research group, Purdue University (2010)

clear all; clc
global hd alpha I_H_conv1 I_H_conv2 Ic
Ic=5.5e-3; %Current at the injector from current source

%%%  LLG parameters
%%%%%%%%%%%%%%%%%%%%%%%%

%%% Constants
%%%------------
q=1.6e-19;                     % Coulombs
hbar=6.626e-34/2/pi;           % Reduced Planck's constant (J-s)
mub=9.274e-21;                 % Bohr Magneton
alpha = 0.007;                 % Gilbert damping parameter
g = 1.76e7;                    % Gyromagnetic ratio [(rad)/(Oe.s)]

%%% Magnet Parameters (taken from experiment)
%%%-------------------------------------------
Ms = 780;                      % Saturation Magnetization [emu/cm^3]
Ku2 = 3.14e4;                  % Uni. anisotropy constant [erg/cm^3]
V1 = (170*75*20)*1e-21;        % Volume [cm^3]
V2 = (170*80*4)*1e-21;         % Volume [cm^3]
Hk = 2*Ku2/Ms   ;              % Switching field [Oe]
Hd = 4*pi*Ms;                  % Demagnetizing field [Oe]
Ns = Ms*V2/mub                 % Number of spins in the magnet

%%% Converting magnet parameters into dimensionless quantities. Note that
%%% in this code we transform the LLG equation into a dimensionless
%%% equation by normalizing it to the time constant 1/(g*Hk).
hk = 1;                    % dimensionless uniaxial field
hd = Hd/Hk;                % dimensionless demag field
tau_c = (1+alpha^2)/(g*Hk); % LLG time constant

% Conversion factor for Ampere spin current into dimensionless input in
% LLG. The factor below is the simplified version of the term
% Is/(q*Ns*g*Hk), noting that g=2muB/hbar.
I_H_conv1 = hbar/2/q/(Ms*V1*Hk*1e-7);
I_H_conv2 = hbar/2/q/(Ms*V2*Hk*1e-7);
Isc = alpha*(1+hd/2)/I_H_conv2;
% Isc =  Estimated ampere spin current required for easy axis switching

%%% Initial conditions of the simulation
mz1=1;
m01=[-sqrt(1-mz1^2) 0 mz1]; %Injector magnet
```

```
mz2=0.99; % Detector magnet slightly off easy axis due to, say, thermal noise
m02=[sqrt(1-mz2^2) 0 mz2]; %Magnet in the x-z plane

%%% Charge current: solving for a fixed number of input current values
%%% since we already have an idea of where switching will occur approximately
Icc=[-8 -5.6 -5.4 -5.3 -4.9 -4.5 -3 -1 1 3 4.5 4.9 5.1 5.4 5.6 8]*1e-3; Nd=length(Icc);

%%%%%%%%%%% Solving the LLG equation
options = odeset('RelTol',1e-8,'AbsTol',1e-9);
NanoS = 50;                    %% Duration in units of nano-seconds
t_span= [0 NanoS*1e-9]/tau_c; %% Dimensionless time span
[t,x]= ode113('LLGsolver', t_span, [m01 m02], options);

mdet_f=zeros(1,Nd);
mdet_r=zeros(1,Nd);

for count=1:Nd
%%%Forward sweep of current
Ic=Icc(count) % Injector current in Amps
[t,x]= ode113('LLGsolver', t_span, [m01 m02], options);
sz=size(t,1);
mdet_f(count)=x(sz,6) %forward sweep
%mz2=mdet_f(count);
%m02=[sqrt(1-mz2^2) 0 mz2];
[Rnl_f(count)]=SpinCircuit(x(sz,1:3), x(sz,4:6));
end

mz2=-0.99; % Detector magnet slightly off easy axis due to, say, thermal noise
m02=[sqrt(1-mz2^2) 0 mz2]; %Magnet in the x-z plane

for count=1:Nd
%%%Reverse sweep of current
Ic=Icc(Nd-count+1) % Injector current in Amps
[t,x]= ode113('LLGsolver', t_span, [m01 m02], options);
sz=size(t,1);
mdet_r(Nd-count+1)=x(sz,6) %reverse sweep
%mz2=mdet_r(count);
%m02=[sqrt(1-mz2^2) 0 mz2];
[Rnl_r(Nd-count+1)]=SpinCircuit(x(sz,1:3), x(sz,4:6));
end

%%%%%%%%%%% Plotting
figure(1) %Non-local resistance v.s. injector charge current
hold on
plot(Icc,Rnl_f*1e3,'b','Linewidth',2); %forward sweep
plot(Icc,Rnl_r*1e3,'r--','Linewidth',2); %reverse sweep
set(gca,'linewidth',3.0,'Fontsize',30)
ylabel('R_{15}(m\Omega)') % The non local V/I
xlabel('I_c (Amp)')
box on
```

```
figure(2) %Magnetization v.s. injector charge current
hold on
plot(Icc,mdet_f,'b','Linewidth',2); %forward sweep
plot(Icc,mdet_r,'r--','Linewidth',2); %reverse sweep
set(gca,'linewidth',3.0,'Fontsize',30)
ylabel('m_z')
xlabel('I_c (Amp)')
box on
%%%%%%%%%%%%%%%%%%%%%%%%%%%%%%%%%%%%%%%%


%%%%%%%%%%%%%%%%%%%%%%%%%%%%%%%%%%%%%%%%
function dmdt = LLGsolver(t,m)
% Vinh Diep, Srikant Srinivasan, Deepanjan Datta, Supriyo Datta Research group
global hd alpha I_H_conv1 I_H_conv2
m1=m(1:3); m2=m(4:6);
H1=[0*m1(1) -hd*m1(2) m1(3)];% Internal fields i.e. uniaxial (along z) and demag(along x)
H2=[0*m2(1) -hd*m2(2) m2(3)];
[Rn1, Is1, Is2]=SpinCircuit(m1, m2);

%%% converting back to [x y z] basis
Is1=Is1([end-1 end end-2])*I_H_conv1; Is1=Is1';
Is2=Is2([end-1 end end-2])*I_H_conv2; Is2=Is2';

%%% Differential Equation for magnetization Dynamics
dm1dt=(-cross(m1,H1)-alpha*cross(m1,cross(m1,H1))...
    +cross(m1,cross(Is1,m1))+alpha*cross(m1,Is1));
dm2dt=(-cross(m2,H2)-alpha*cross(m2,cross(m2,H2))...
    +cross(m2,cross(Is2,m2))+alpha*cross(m2,Is2));
dmdt=[dm1dt'; dm2dt'];

end
%%%%%%%%%%%%%%%%%%%%%%%%%%%%%%%%%%%%%%%%


%%%%%%%%%%%%%%%%%%%%%%%%%%%%%%%%%%%%%%%%
function [Rnl, Is1, Is2]=SpinCircuit(m1,m2)
% 4-component Spin Circuit for the device in ref [Otani].
% Srikant Srinivasan, Purdue University Sept. 28, 2010
global Ic

zdir=[1 0 0]; % Unit vector along 'z', the basis convention being [z x y]
m1=m1([end 1:end-1]);
m2=m2([end 1:end-1]);

% Constants (all MKS)
q=1.6e-19; h=6.626e-34;
Z=zeros(4);
%%%%%%% Expt. Ckt. Parameters (SI units)
%%%%%%%%%%%%%%%%%%%%%%%%%%%%%%%%%%%%%%%%%%
% Magnet
```

```
PF1=0.49;PF2=0.49; %Magnet and Interface polarizations
AF1=170*75e-18; lF1=20e-9;   % Area, thickness of Magnet 1
AF2=80*170e-18; lF2=4e-9;   % Area, thickness of Magnet 2
lambdaF=5e-9; rhoF=17.1e-8;  % Permalloy resistivity and spin-flip length
RF1=lambdaF*rhoF/AF1; RF2=lambdaF*rhoF/AF2; %Parameters of magnets
LF1=lF1/lambdaF; LF2=lF2/lambdaF;  % Normalized magnet length
kf=1.36e10; Modes=kf*kf/2/pi; % Modes including both spins
RqF=h/q/q; % quantum of resistance per spin

% Channel
t=65e-9; AN=170e-9*t;   % thickness, cross sectional area of Channel
lambdaN=1e-6; rhoN=0.69e-8; RN=lambdaN*rhoN/AN; % Copper
RN1=RN; RN2=RN; RN3=RN; % RN2=channel between inj. and det. and RN1,3=overhanging regions
lN2=270e-9; LN2=lN2/lambdaN; LN1=10; LN3=10;

% Gold lead
lambdaG=1e-8;rhoG=7e-8;Rau=lambdaG*rhoG/AF1;Lau=10;

%%%%% Spin Ckt Description
%%%%%%%%%%%%%%%%%%%%%%%%%%%%%%
% Conductances

% Non-magnetic channel
[GN1,G0N1]=G_4x4(RN1,LN1,0,0,0);
[GN2,G0N2]=G_4x4(RN2,LN2,0,0,0);
[GN3,G0N3]=G_4x4(RN3,LN3,0,0,0);

% Top Gold Contacts
[GA1,G0A1]=G_4x4(Rau,Lau,0,0,0);
[GA2,G0A2]=G_4x4(Rau,Lau,0,0,0);

% Ferromagnet Bulk
[GF1,G0F1]=G_4x4(RF1,LF1,PF1,0,0);
[GF2,G0F2]=G_4x4(RF2,LF2,PF2,0,0);

% Ferromagnetic Interfaces
[GBF1,G0BF1]=G_4x4(RqF/(Modes*AF1),0,PF1,1,0);
[GBF2,G0BF2]=G_4x4(RqF/(Modes*AF2),0,PF2,1,0);
G0F1=G0F1+G0BF1;G0F2=G0F2+G0BF2;
% if max(max(GBF1))<max(max(GF1))
%     % Ballistic limit
%     GF1=GBF1; G0F1=G0BF1;
% else
%     % diffusive limit
%     G0F1=G0F1+G0BF1;
% end
% if max(max(GBF2))<max(max(GF2))
%     % Ballistic limit
%     GF2=GBF2; G0F2=G0BF2;
% else
%     % diffusive limit
```

```
%    G0F2=G0F2+G0BF2;
% end

U1=rotmat(zdir,m1);
GF1=U1*GF1*U1'; G0F1=U1*G0F1*U1';
U2=rotmat(zdir,m2);
GF2=U2*GF2*U2'; G0F2=U2*G0F2*U2';

% Non-local computation
% Conductance matrix
G=[G0A1+GA1 -GA1 Z Z Z Z Z;
   -GA1 G0A1+GA1+G0F1+GF1 -GF1 Z Z Z Z;
   Z -GF1 G0F1+GF1+GN2+G0N2+G0N1+GN1 -GN2 Z Z Z ;
   Z Z -GN2 G0N2+GN2+G0F2+GF2+GN3+G0N3 -GF2 Z -GN3;
   Z Z Z -GF2 GF2+G0F2+GA2+G0A2 -GA2 Z;
   Z Z Z Z -GA2 GA2+G0A2 Z;
   Z Z Z -GN3 Z Z GN3+G0N3];

C = [Ic;zeros(27,1)];%  Terminal currents
V=G\C;V=reshape(V,4,7);% Terminal voltages
delV=V(1,6)-V(1,7); % Non-Local voltage measured

Rnl=delV/Ic; % Non-Local Resistance
IF1=GF1*(V(:,3)-V(:,2))+(G0F1)*V(:,3);% current entering FM1
IF2=GF2*(V(:,4)-V(:,5))+(G0F2)*V(:,4);% current entering FM2
Is1=-IF1(2:4); Is2=-IF2(2:4); % Electron spin current
end

%%%%%%%%%%%%%%%%%%%%%%%%%%%%%%%%%%%%%%%%%%%%%%%%%%%%%%%%%%%%%%%%%%%%%%%%%%%%%%%%%%%%%%%%


%%%%%%%%%%%%%%%%%%%%%%%%%%%%%%%%%%%%%%%%%%%%%%%%%%%%%%%%%%%%%%%%%%%%%%%%%%%%%%%%%%%%%%%%
function [Gmat, G0mat] = G_4x4(R,L,P,eta,ang)
% This function generates a 4 component conductance matrix for the various
% sections including Ferro-magnet, tunnel barrier, non-magnetic channel,
% Interface.
% Srikant Srinivasan, Purdue University (2010)

% Inputs of this function are defined for each as:
%--------------------------------------------------
% R=Spin resistance (i.e rho*lambda/A),
% L=Length normalized to spin diffusion length i.e. L/lambda,
% P=polarization fraction in the range (-1,1),
% eta=ratio of (mixing conductance/series conductance),
% ang=mixing angle (Ratio of Slonczewski:Field-Like Spin torque).
a=cos(ang);b=sin(ang);

% Setting up the Series (Gmat) and Shunt (G0mat) conductance matrices
%-------------------------------------------------------------------
Gmat=[1 P 0 0; P P^2 0 0; 0 0 0 0; 0 0 0 0];
```

```
if eta==0
   % Individual Sections (eta>0 is defined for the interface)
   if L>0
      if P==0
         % Non Magnet
         Gmat=(1/R/L)*(Gmat+ L*csch(L)*diag([0 1 1 1]));
         G0mat=1/R*tanh(L/2)*diag([0 1 1 1]);
      else
         % Ferro Magnet
         Gmat=(1/R/L)*(Gmat+(1-P^2)*L*csch(L)*diag([0 1 0 0]));
         G0mat=(1-P^2)/R*tanh(L/2)*diag([0 1 0 0]);
      end
   else
      %tunnel barrier (heuristic extension of 2 component)
      Gmat=(1/R)*[1 P 0 0; P 1 0 0; 0 0 1 0; 0 0 0 1];
      G0mat=[];
   end
else
   % FM/NM Interface conductance (based on derivation in Appendix B)
   Gmat=1/eta/R*[1 P 0 0; P 1 0 0; 0 0 0 0; 0 0 0 0];
   G0mat=1/R*[0 0 0 0; 0 0 0 0; 0 0 a b; 0 0 -b a];
end
end
%%%%%%%%%%%%%%%%%%%%%%%%%%%%%%%%%%%%%%%


%%%%%%%%%%%%%%%%%%%%%%%%%%%%%%%%%%%%%%%
function R=rotmat(a,b)
% Implementing Rodriguez rotation formula to transform the conductance
% matrix for a magnet aligned along a direction specified by the vector 'a'
% to a direction specified by the vector 'b'
% Srikant Srinivasan, Purdue University (2010)

a=a/norm(a); b=b/norm(b);
c=dot(a,b); s=sqrt(1-c^2);
if s==0
   % Initial and final vectors are collinear
   u=[0 0 0];
else
   u=cross(a,b)/norm(cross(a,b));
end

%%% Z,X,Y coordinate system
ux=u(2); uy=u(3); uz=u(1);
R=[1      0                  0                  0;
   0 uz^2+(1-uz^2)*c      uz*ux*(1-c)-uy*s    uz*uy*(1-c)+ux*s;
   0 uz*ux*(1-c)+uy*s     ux^2+(1-ux^2)*c     ux*uy*(1-c)-uz*s;
   0 uz*uy*(1-c)-ux*s     ux*uy*(1-c)+uz*s    uy^2+(1-uy^2)*c];
end

%%%%%%%%%%%%%%%%%%%%%%%%%%%%%%%%%%%%%%%
```